\documentclass[%
 reprint,
 superscriptaddress,
 showpacs,preprintnumbers,
 nofootinbib,
 amsmath,amssymb,
 aps,
]{revtex4-1}

\usepackage{graphicx}
\graphicspath{{fig/}}   
\usepackage{dcolumn}
\usepackage{bm}
\usepackage{makecell}
\usepackage{braket} 
\usepackage{siunitx}
\usepackage{color}
\usepackage{xr}

\newcommand*{\I}{\mathrm{i}}

\begin{document}

\title{Analog quantum simulation of the Rabi model in the ultra-strong coupling regime}

\author{Jochen Braum\"uller}
\affiliation{Physikalisches Institut, Karlsruhe Institute of Technology, Wolfgang-Gaede-Str. 1, 76131 Karlsruhe, Germany}

\author{Michael Marthaler}
\affiliation{Institut f\"ur Theoretische Festk\"orperphysik, Karlsruhe Institute of Technology, Wolfgang-Gaede-Str. 1, 76131 Karlsruhe, Germany}

\author{Andre Schneider}
\author{Alexander Stehli}
\author{Hannes Rotzinger}
\affiliation{Physikalisches Institut, Karlsruhe Institute of Technology, Wolfgang-Gaede-Str. 1, 76131 Karlsruhe, Germany}

\author{Martin Weides}
\affiliation{Physikalisches Institut, Karlsruhe Institute of Technology, Wolfgang-Gaede-Str. 1, 76131 Karlsruhe, Germany}
\affiliation{Physikalisches Institut, Johannes Gutenberg University Mainz, Staudinger Weg 7, 55128 Mainz, Germany}

\author{Alexey V. Ustinov}
\affiliation{Physikalisches Institut, Karlsruhe Institute of Technology, Wolfgang-Gaede-Str. 1, 76131 Karlsruhe, Germany}
\affiliation{Russian Quantum Center, National University of Science and Technology MISIS, Leninsky Ave 4, Moscow 119049, Russia}

\date{\today}

\begin{abstract}
The quantum Rabi model describes the fundamental mechanism of light-matter interaction. It consists of a two-level atom or qubit coupled to a quantized harmonic mode via a transversal interaction. In the weak coupling regime, it reduces to the well-known Jaynes-Cummings model by applying a rotating wave approximation (RWA). The RWA breaks down in the ultra-strong coupling (USC) regime, where the effective coupling strength $g$ is comparable to the energy $\omega$ of the bosonic mode, and remarkable features in the system dynamics are revealed. We demonstrate an analog quantum simulation of an effective quantum Rabi model in the USC regime, achieving a relative coupling ratio of $g/\omega \sim 0.6$. The quantum hardware of the simulator is a superconducting circuit embedded in a cQED setup. We observe fast and periodic quantum state collapses and revivals of the initial qubit state, being the most distinct signature of the synthesized model.
\end{abstract}


\maketitle

Finding solutions to many quantum problems is a very challenging task \cite{Feynman1982}. The reason is the exponentially large number of degrees of freedom in a quantum system, requiring computational power and memory that easily exceed the capabilities of present classical computers. A yet to be demonstrated universal digital quantum computer of sufficient size would be capable of efficiently solving most quantum problems \cite{Feynman1982,Lloyd1996}. A more feasible approach to achieve a computational speedup in the near future is quantum simulation \cite{Feynman1982,Lloyd1996,Georgescu2014}. In the framework of analog quantum simulation, a tailored and well-controllable artificial quantum system is mapped onto a quantum problem of interest in order to mimic its dynamics. Since the same equations of motion hold for both systems, the solution of the underlying quantum problem is inferred by observing the time evolution of the artificially built model system while making use of its intrinsic quantumness. This scheme may be applied to the simulation of complex quantum problems, in the spirit originally proposed by Feynman \cite{Feynman1982}.

Quantum simulation was performed on various experimental platforms. Examples of analog quantum simulation are the study of fermionic transport \cite{Schneider2012} and magnetism \cite{Greif2013} with cold atoms and the simulation of a quantum magnet and the Dirac equation with trapped ions \cite{Friedenauer2008,Gerritsma2010}. The exploration of non-equilibrium physics was proposed with an on-chip quantum simulator based on superconducting circuits \cite{Houck2012,Reiner2016}. Digital simulation schemes with superconducting devices were demonstrated for fermionic models \cite{Barends2015} and spin systems \cite{Salathe2015}.

The quantum Rabi model in quantum optics describes the interaction between a two-level atom and a single quantized harmonic oscillator mode \cite{Rabi1936,Rabi1937}. In the weak coupling regime, which may still be strong in the sense of quantum electrodynamics (QED), a RWA can be applied and the Rabi model reduces to the Jaynes-Cummings model \cite{Jaynes1963}, which captures most relevant scenarios in cavity and circuit QED. In the USC and deep strong coupling regimes, where the coupling strength is comparable to the mode energies \cite{Casanova2010}, the counter rotating terms in the interaction Hamiltonian can no longer be neglected and the RWA breaks down. As a consequence, the total excitation number in the quantum Rabi model is not conserved. Except for one recent paradigm of finding an exact solution \cite{Braak2011}, an analytically closed solution of the quantum Rabi model does not exist due to the lack of a second conserved quantity which renders it non-integrable. The quantum Rabi model, in particular in the USC regime and beyond, exhibits non-classical features and rising interest in it is inspired by strong advances of experimental capabilities \cite{Casanova2010,Ashhab2010,Beaudoin2011,Alderete2016}. The specific spectral features of the USC regime and the consequent breakdown of the RWA were previously observed with a superconducting circuit by implementing an increased physical coupling strength \cite{Niemczyk2010,Forn-DiazRom2016}. A similar approach involving a flux qubit coupled to a single-mode resonator allowed to access the deep strong coupling regime in a closed system \cite{Yoshihara2016}. The USC regime was reached before by dynamically modulating the flux bias of a superconducting qubit, reaching a coupling strength of about $0.1$ of the effective resonator frequency \cite{Li2013}.

In our approach, we engineer an effective quantum Rabi Hamiltonian with an analog quantum simulation scheme based on the application of microwave Rabi drive tones. By a decrease of the subsystem energies, the USC condition is satisfied in the effective rotating frame, allowing to observe the distinct model dynamics. The scheme may be a route to efficiently generate non-classical cavity states \cite{Leghtas2015,Kirchmair2013,Vlastakis2013} and may be extended to explore relevant physical models such as the Dirac equation in (1+1) dimensions. Its characteristic dynamics is expected to display a \textit{Zitterbewegung} in the spacial quadrature of the bosonic mode \cite{Ballester2012}. This dynamics has been observed with trapped ions \cite{Gerritsma2010}, likewise based on a Hamiltonian that is closely related to the USC Rabi model. It has been shown recently that a quantum phase transition, typically requiring a continuum of modes, can appear already in the quantum Rabi model under appropriate conditions \cite{Hwang2015}. The experimental challenge is projected to the coupling requirements in the model which may be accomplished with the simulation scheme presented. This can be a starting point to experimentally investigate critical phenomena in a small and well-controlled quantum system \cite{Hwang2016}. With a digital simulation approach, the dynamics of the quantum Rabi model in USC conditions was similarly studied very recently \cite{Langford2016}.

In our experiment we simulate the quantum Rabi model in the USC regime achieving a relative coupling strength of up to $0.6$. Dependent on our experimental parameters, we observe periodically recurring quantum state collapses and revivals in the qubit dynamics, being a distinct signature of USC. The collapse-revival dynamics appears most clearly in the absence of the qubit energy term in the model, according to the expectation from master equation simulations. In addition, we use our device to simulate the full quantum Rabi model and are able to observe the onset of an additional substructure in the qubit time evolution. With this proof of principle experiment we validate the experimental feasibility of the analog quantum simulation scheme and demonstrate the potential of superconducting circuits for the field of quantum simulation.

\section*{Results}
\subsection*{Simulation scheme}
The quantum Rabi Hamiltonian reads
\begin{equation}
\frac{\hat H}{\hbar} =\frac{\epsilon} 2 \hat\sigma _z+\omega\hat b ^\dagger \hat b+g\hat\sigma _x\left(\hat b ^\dagger +\hat b\right),
\label{eq:HRabi}
\end{equation}
with $\epsilon$ the qubit energy splitting, $\omega$ the bosonic mode frequency and $g$ the transversal coupling strength. $\hat\sigma _i$ are Pauli matrices with $\hat\sigma _z\Ket{\mathrm{g}}=-\Ket{\mathrm{g}}$ and $\hat\sigma _z\Ket{\mathrm{e}}=\Ket{\mathrm{e}}$, where $\Ket{\mathrm{g}}$, $\Ket{\mathrm{e}}$ denote eigenstates of the computational qubit basis. $\hat b ^\dagger$ ($\hat b$) are creation (annihilation) operators in the Fock space of the bosonic mode. Both elements of the model are physically implemented in the experiment, with a small geometric coupling $g\ll\epsilon,\,\omega$, such that the RWA applies and Eq.~\eqref{eq:HRabi} takes the form of the Jaynes-Cummings Hamiltonian. In order to access the USC regime, we follow the scheme proposed in Ref.~\cite{Ballester2012}. It is based on the application of two transversal microwave Rabi drive tones coupling to the qubit. The USC condition is created in a synthesized effective Hamiltonian in the frame rotating with the dominant drive frequency. In this engineered Hamiltonian, the effective mode energies are set by the Rabi drive parameters. The Jaynes-Cummings Hamiltonian in the laboratory frame with both drives applied takes the form
\begin{align}
\frac{\hat H _{\mathrm{d}}}{\hbar} =&\frac{\epsilon} 2 \hat\sigma _z+\omega\hat b ^\dagger \hat b +g\left(\hat\sigma _-\hat b ^\dagger +\hat\sigma _+\hat b\right)\nonumber\\
&+\hat\sigma _x \eta_1\cos(\omega_1 t+\varphi_1)+\hat\sigma _x \eta_2\cos(\omega_2 t+\varphi_2),
\label{eq:HDrives}
\end{align}
with $\eta_i$ the amplitudes and $\omega_i$ the frequencies of drive $i$. $\varphi_i$ denotes the relative phase of drive $i$ in the coordinate system of the qubit Bloch sphere in the laboratory frame. Within the RWA where $\eta_i/\omega_i\ll 1$, the $\varphi_i$ enter as relative phases of the transversal coupling operators $e^{-\I\varphi_i}\hat\sigma _+ +\mathrm{h.c.}$, where $\hat\sigma_{\pm} =1/2\left(\hat\sigma _x\pm \I\hat\sigma _y\right)$ denote Pauli's ladder operators. In the following, we set $\varphi_i=0$ to recover the familiar $\hat\sigma _x$ coupling without loss of generality. Going to the frame rotating with $\omega_1$ and neglecting terms rotating with $e^{\pm 2\I\omega_1 t}$ renders the first driving term time-independent, yielding
\begin{align}
\frac{\hat H _1}{\hbar} =& \left(\epsilon-\omega_1\right)\frac{\hat\sigma _z}2 +\left(\omega-\omega_1\right)\hat b ^\dagger \hat b +g\left(\hat\sigma _-\hat b ^\dagger +\hat\sigma _+\hat b\right)\nonumber\\
&+\frac{\eta_{1}}{2}\hat\sigma _x +\frac{\eta_2}2\left(\hat\sigma _+ e^{\I(\omega_1-\omega_2)t}+\hat\sigma _- e^{-\I(\omega_1-\omega_2)t}\right).
\label{eq:Hrot1}
\end{align}
The $\eta_1$-term is now the significant term and we move into its interaction picture. Satisfying the requirement $\omega_1-\omega_2 =\eta_1$ and applying a RWA yields the effective Hamiltonian in the $\omega_1$ frame
\begin{equation}
\frac{\hat H _{\mathrm{eff}}}{\hbar} = \frac{\eta_2}2\frac{\hat\sigma _z}2 +\omega_{\mathrm{eff}}\hat b ^\dagger \hat b +\frac g 2\hat\sigma _x\left(\hat b ^\dagger +\hat b\right).
\label{eq:Hrot1eff}
\end{equation}
We define the effective bosonic mode energy $\omega_{\mathrm{eff}}\equiv \omega-\omega_1$, which is the parameter governing the system dynamics. Noting $\eta_1\gg \eta_2$, which is a necessary condition for the above approximation to hold, the effective qubit frequency $\eta_2$ and effective bosonic mode frequency $\omega_{\mathrm{eff}}$ can be chosen as experimental parameters in the simulation. The complete coupling term of the quantum Rabi Hamiltonian is recovered, valid in the USC regime and beyond, while the geometric coupling strength is only modified by a factor of two, resulting in $g_{\mathrm{eff}}=g/2$. It is therewith feasible to tune the system into a regime where the coupling strength is similar to or exceeds the subsystem energies. This is achieved by leaving the geometric coupling strength essentially unchanged in the synthesized Hamiltonian, while slowing down the system dynamics by effectively decreasing the mode frequencies to $\lesssim \SI{8}{MHz}$. Thermal excitations of these effective transitions can be neglected since they couple to the thermal bath excitation frequency $\sim \SI{1}{GHz}$ of the cryostat via their laboratory frame equivalent frequency of $\omega_1/2\pi\sim\SI{6}{GHz}$. We want to point out that the coupling regime is defined by $g_{\mathrm{eff}}/\omega_{\mathrm{eff}}$, rather than involving the Rabi frequency $\eta_1$, which does not enter the synthesized Hamiltonian. While the simulation scheme requires $|\epsilon -\omega_1|\ll\eta_1$, the qubit frequency does not enter the effective Hamiltonian. The time evolution of the qubit measured in the laboratory frame is subject to fast oscillations corresponding to the Rabi frequency $\eta_1$. Accordingly, the qubit dynamics in the engineered quantum Rabi Hamiltonian Eq.~\eqref{eq:Hrot1eff}, valid in the $\omega_1$ frame, can be inferred from the envelope of the evolution in the laboratory frame. The derivation of Eq.~\eqref{eq:Hrot1eff} can be found in Ref.~\cite{Ballester2012} and is detailed in Supplementary Note 1. A similar drive scheme based on a Rabi tone was previously used in experiment to synthesize an effective Hamiltonian with a rotated qubit basis \cite{Vool2016}. For the qubit and the bosonic mode degenerate in the laboratory frame, a distinct collapse-revival signature appears in the dynamics of the quantum Rabi model under USC conditions.

\subsection*{Quantum simulation device}
\begin{figure} 
\includegraphics{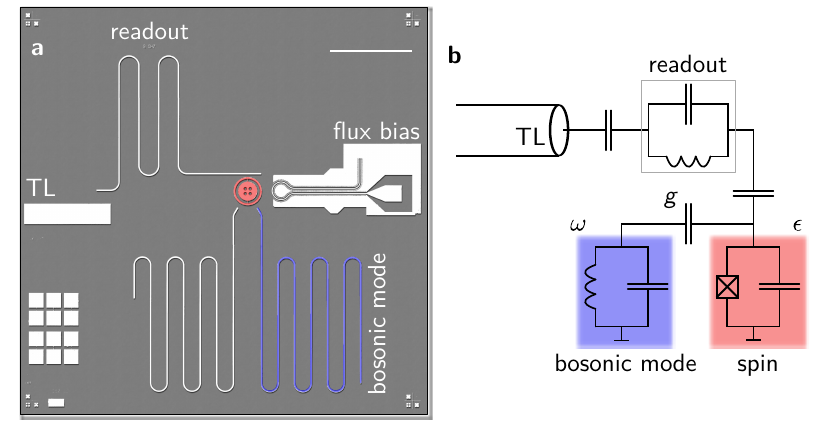}
\caption{\textbf{Quantum simulation device} (a) Optical micrograph with the atomic spin represented by a concentric transmon qubit, highlighted in red and the $\lambda/2$ microstrip resonator (blue) constituting the bosonic oscillator mode. The readout resonator couples to the qubit capacitively and is read out with an open transmission line (TL) via the reflection signal of an applied microwave tone or pulse. The second resonator visible on chip is not used in the current experiment and is detuned in frequency from the relevant bosonic mode by $\sim\SI{0.5}{GHz}$. The scale bar corresponds to $\SI{1}{mm}$. (b) Effective circuit diagram of the device.} 
\label{fig:chip_cross}
\end{figure}

The physical implementation of the quantum simulator is based on a superconducting circuit embedded in a typical circuit QED setup \cite{Blais2004,Wallraff2004}, see Fig.~\ref{fig:chip_cross}. The atomic spin of the quantum Rabi model is mapped to a concentric transmon qubit \cite{Koch2007,Braumueller2016}. It is operated at a ratio of Josephson energy to charging energy $E_{\mathrm{J}}/E_{\mathrm{C}}= 50$ and an anharmonicity $\alpha/h = \omega_{12}/2\pi -\omega_{01}/2\pi = \SI{-0.36}{GHz}\sim -E_{\mathrm{C}}/h=-\SI{0.31}{GHz}$, close to resonance with the bosonic mode at $\SI{5.948}{GHz}$. $\omega_{ij}$ denote the transition frequencies between transmon levels $i$, $j$. The energy relaxation rate of the qubit at the operation point is measured to be $1/T_1=0.2\times 10^ 6\,\mathrm{s^{-1}}$. An on-chip flux bias line allows for a fast tuning of the qubit transition frequency as the concentric transmon is formed by a gradiometric dc SQUID. The bosonic mode of the model is represented by a harmonic $\lambda/2$ resonator with an inverse lifetime $\kappa\sim 3.9\times 10^ 6\,\mathrm{s^{-1}}$ that is limited by internal loss. Following the common convention, we use $\kappa$ as the inverse photon lifetime of a linear cavity, which may be extracted as the full width at half maximum of a resonance signature in frequency space. Via Fourier transformation one can see that this means the cavity relaxes to its groundstate at a rate of $\kappa /2$. In a separate experiment we find the internal quality of similar microstrip resonators to be limited to about $1.2\times 10^ 4$ in the single photon regime, corresponding to a loss rate of $3.1\times 10^ 6\,\mathrm{s^{-1}}$. Microwave simulations indicate that the quality is limited by radiation. The sample fabrication process is detailed in Supplementary Note 2.

\subsection*{Sample characterization}
\begin{figure} 
\includegraphics{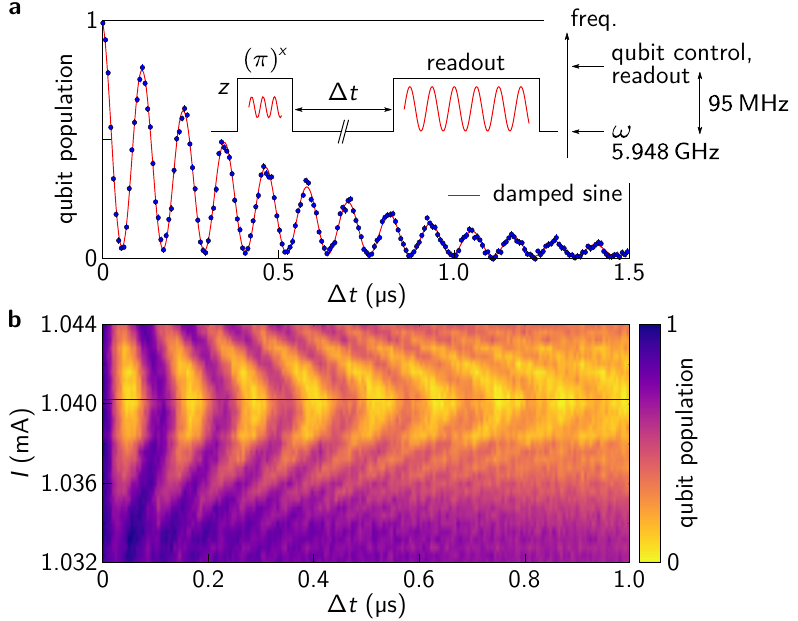}
\caption{\textbf{Vacuum Rabi oscillations between qubit and bosonic mode} (a) The qubit is initially dc-biased on resonance with the bosonic mode, while it is detuned for state preparation and readout. The solid black line in the inset depicts the fast flux pulses applied to the flux bias line and indicates the qubit frequency on the given axis. Qubit and bosonic mode are on resonance during an interaction time $\Delta t$. A frequency fit (red) of the vacuum Rabi oscillations yields $2g/2\pi=\SI{8.5}{MHz}$. With the decay rate $\Gamma=(2.08\pm 0.03)\times 10^ 6\,\mathrm{s^{-1}}$ of the envelope and the qubit decay rate $1/T_1=(0.2\pm 0.12)\times 10^ 6\,\mathrm{s^{-1}}$ we extract the bosonic mode decay rate $\kappa =(3.9\pm 0.13)\times 10^ 6\,\mathrm{s^{-1}}$. Error bars denote a statistical standard deviation as detailed in the Methods. (b) For departing from the resonance condition (blue line) by varying the dc bias current $I$, we observe the expected decrease in excitation swap efficiency and an increase in the vacuum Rabi frequency. The qubit population is given in colors and we applied a numerical interpolation of data points.}
\label{fig:vacRabi}
\end{figure}

The quantum state collapse followed by a quantum revival is the most striking signature of the ultra-strong and close deep strong coupling regime of the quantum Rabi model and emerges for qubit and bosonic mode being degenerate in the laboratory frame. We calibrate this resonance condition by minimizing the periodic swap rate of a single excitation between qubit and bosonic mode for the simple Jaynes-Cummings model in the absence of additional Rabi drives. Figure~\ref{fig:vacRabi} shows the measured vacuum Rabi fluctuations in the resonant case (a) and dependent on the qubit transition frequency (b). For initial state preparation of the qubit and readout we detune the qubit by $\SI{95}{MHz}$ to a higher frequency. This corresponds to switching off the resonant interaction with the bosonic mode. Supplementary Note 3 describes experimental details on flux pulse generation. Rabi vacuum oscillations can be observed during the interaction time $\Delta t$ and yield a coupling strength $g/2\pi=\SI{4.3}{MHz}$, in good agreement with the spectroscopically obtained result, see Supplementary Note 8.

\subsection*{Quantum state collapse and revival}
\begin{figure*} 
\includegraphics{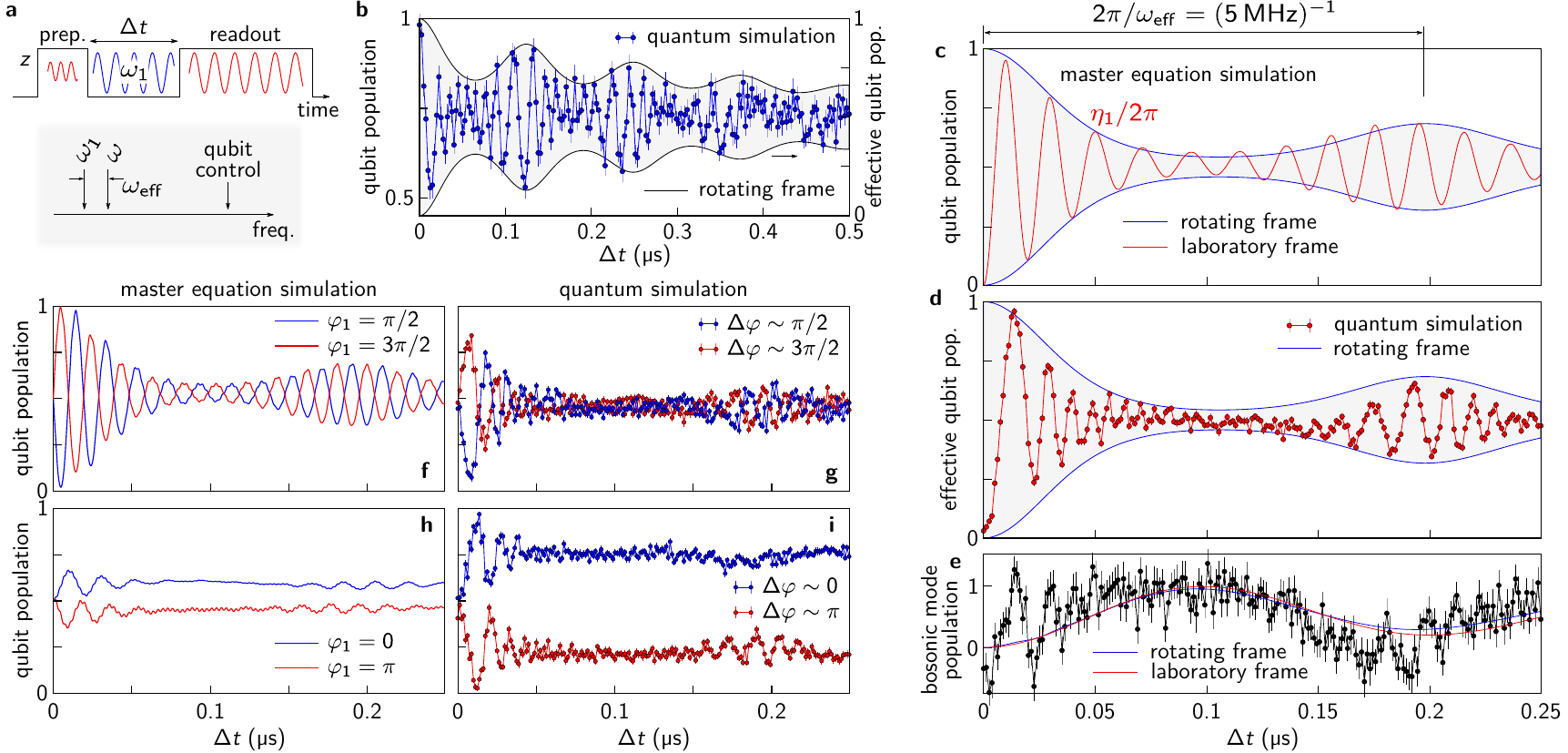}
\caption{\textbf{Quantum state collapse and revival with only the dominant Rabi drive applied} (a) Schematic pulse sequence and overview on the relative frequencies used in the experiment. (b) Quantum simulation of the periodic recurrence of quantum state revivals for $\omega_{\mathrm{eff}}/2\pi=\SI{8}{MHz}$. The blue line corresponds to a master equation simulation of the qubit evolution in the rotating frame. (c), (d) Master equation and quantum simulation of the qubit time evolution for initial qubit states $\Ket{\mathrm{g}}$, $\Ket{\mathrm{e}}$ and $\omega_{\mathrm{eff}}/2\pi=\SI{5}{MHz}$, corresponding to $g_{\mathrm{eff}}/\omega_{\mathrm{eff}}\sim 0.5$. The red line shows the qubit population evolution of the driven system in the laboratory frame, Eq.~\eqref{eq:HDrives}, while the blue lines follow the qubit evolution in the synthesized Hamiltonian Eq.~\eqref{eq:Hrot1eff}, likewise extracted from a classical master equation simulation. The deviation between the envelope of the laboratory frame data and the rotating frame data in (c) reflects the approximations of the simulation scheme. Experimental data shows the difference between two measurements for the qubit prepared in $\Ket{\mathrm{g}}$, $\Ket{\mathrm{e}}$, respectively, in order to isolate the qubit signal. (e) Measured population evolution of the bosonic mode, extracted from the sum of the two successive measurements and fitted to classically simulated data. (f)-(i) Qubit time evolution for varying relative phase $\varphi_1$ of the applied drive. The initial qubit state is prepared on the equator of the Bloch sphere $\Ket{\mathrm{g}}\pm\Ket{\mathrm{e}}$. Dispersive shifts induced by the bosonic mode are subtracted based on its classically simulated population evolution. Error bars throughout the figure denote a statistical standard deviation as detailed in the Methods.}
\label{fig:revival}
\end{figure*}

As the collapse-revival signature of the quantum Rabi model in USC conditions manifests most clearly for a vanishing qubit term, we initially set $\eta_2=0$, yielding the effective Hamiltonian in the qubit frame
\begin{equation}
\frac{\hat H}{\hbar} = \omega_{\mathrm{eff}}\hat b ^\dagger \hat b +\frac g 2\hat\sigma _x\left(\hat b ^\dagger +\hat b\right).
\label{eq:quframe}
\end{equation}
Figure~\ref{fig:revival}(a) shows the applied measurement sequence which is based on the one in Fig.~\ref{fig:vacRabi} but extended by a drive tone of amplitude $\eta_1$. The bosonic mode is initially in the vacuum state and the qubit is prepared in one of its basis states $\Ket{\mathrm{g}}$, $\Ket{\mathrm{e}}$, which are thermally impure. Qubit and bosonic mode are on resonance during the simulation time $\Delta t$. The drive is applied at a frequency $\omega_1$ detuned from the common resonance point by $\omega_{\mathrm{eff}}$, setting the effective bosonic mode frequency in the rotating frame. Measured data for $\omega_{\mathrm{eff}}/2\pi=\SI{8}{MHz}$ is displayed in Fig.~\ref{fig:revival}(b), corresponding to $g_{\mathrm{eff}}/\omega_{\mathrm{eff}}\sim 0.3$. Data points show the experimentally simulated time evolution of the qubit prepared in $\Ket{\mathrm{e}}$. A fast quantum state collapse followed by periodically returning quantum revivals can be observed. The ground state of the qubit subspace in the driven system as well as in the synthesized Hamiltonian, Eq.~\eqref{eq:quframe}, is in the equatorial plane of the qubit Bloch sphere and is occupied after a time $\Delta t \gg T_1,\,1/\kappa$. It is diagonal in the $\Ket\pm$ basis, with $\Ket\pm=1/\sqrt 2 (\Ket{\mathrm{e}}\pm\Ket{\mathrm{g}})$. The revival dynamics can be understood with an intuitive picture in the laboratory frame. The eigenenergies in the $\Ket\pm$ subspaces take the form of displaced vacuum
\begin{equation}
\omega_{\mathrm{eff}}\left(\hat b ^\dagger \pm\frac g{2\omega_{\mathrm{eff}}}\right)\left(\hat b \pm\frac g{2\omega_{\mathrm{eff}}}\right) +\mathrm{const.},
\end{equation}
which is a coherent state that is not diagonal in the Fock basis. The prepared initial state in the experiment is therefore not an eigenstate in the effective basis with the drive applied such that many terms corresponding to the relevant Fock states $n$ of the bosonic mode participate in the dynamics with phase factors $\exp\{\I n\omega_{\mathrm{eff}}t\}$, $n\in \mathbb{N}^+$. While contributing terms get out of phase during the state collapse, they rephase after an idling period of $2\pi/\omega_{\mathrm{eff}}$ to form the quantum revival. The underlying physics of this phenomenon is fundamentally different from the origin of state revivals that were proposed for the Jaynes-Cummings model \cite{Eberly1980}. Here, the preparation of the bosonic mode in a large coherent state with $\alpha \gtrsim 10$ is required and non-periodic revivals are expected at times $\propto 1/g_{\mathrm{eff}}$ rather than $\propto 1/\omega_{\mathrm{eff}}$ \cite{Haroche2006}, as demonstrated in Supplementary Figure 7.
The blue line in Fig.~\ref{fig:revival}(b) corresponds to a classical master equation simulation of the qubit dynamics in the rotating frame in the two-level approximation. It includes the second excited level of the transmon \cite{Braumuller2015} and decay terms in the underlying Liouvillian according to measured values. Refer to Supplementary Note 5 for further details. Figures~\ref{fig:revival}(c),(d) show a classical simulation and the quantum simulation for $\omega_{\mathrm{eff}}/2\pi =\SI{5}{MHz}$ with the qubit prepared in one of its eigenstates $\Ket{\mathrm{g}}$, $\Ket{\mathrm{e}}$. The population of the bosonic mode takes a maximum during the idling period and adopts its initial population at $2\pi/\omega_{\mathrm{eff}}$ in the absence of dissipation, see Fig.~\ref{fig:revival}(e). The fast oscillations in Fig.~\ref{fig:revival}(c), (d) correspond to the Rabi frequency $\eta_1/2\pi\sim\SI{50}{MHz}$. This value is chosen such that the requirement $\eta_1/\omega_{\mathrm{eff}}\gg 1$ is fulfilled while staying well below the transmon anharmonicity, avoiding higher level populations. Deviations in the laboratory frame simulation traces are due to a uncertainty in the Rabi frequency that is extracted from Fourier transformation of measured data. The broadening in frequency space is mainly caused by the beating in experimental data, which is an experimental artifact. The relevant dynamics of the USC quantum Rabi Hamiltonian corresponds to the envelope of measured data. Since the laboratory frame dissipation is enhanced for a larger ratio of photon population in the bosonic mode, the accessible coupling regime is bound by the limited coherence of the bosonic mode, in particular. This is reflected in a dependence of the coherence envelope of the quantum revivals on the ratio $g/\omega_{\mathrm{eff}}$, see Supplementary Figure 7, reflecting that the excitation number is no longer a conserved quantity in the quantum Rabi model. We find a better agreement with experimental data for using a slightly increased value for the geometric coupling strength in the master equation simulation than extracted from vacuum Rabi oscillations. See Supplementary Notes 3 and 6 for a discussion and a summary of the relevant parameters.

The validity of the analog simulation scheme proposed in Ref.~\cite{Ballester2012} and used in this letter is confirmed by master equation simulations given in Supplementary Figure 4. For ideal conditions, we demonstrate that the dynamics of the qubit and the bosonic mode in the quantum Rabi model is well reproduced by the constructed effective Hamiltonian and that the population of the bosonic mode is independent of the Rabi drive amplitude $\eta_1$, despite of it forming a large energy reservoir that is provided to the circuit.

In the experiment we face a parasitic coupling of the Rabi tones to the bosonic mode that is degenerate to the qubit and spatially close by in the circuit. This leads to an excess population of the bosonic mode, however without disturbing the functional evolution of its population. This is evident as the evolution of the simple harmonic Hamiltonian $\hat H _{\mathrm{h}}/\hbar = \omega_{\mathrm{eff}}\hat b ^\dagger \hat b +\frac 1 2 \eta_{\mathrm{r}}(\hat b ^\dagger +\hat b)$ agrees with the expectation for the quantum Rabi model up to a scaling factor, where the last term corresponds to the parasitic drive of strength $\eta_{\mathrm{r}}$ transformed to the rotating frame. By performing the displacement transformation $\hat D =\exp{\{-\eta_{\mathrm{r}}/(2\omega_{\mathrm{eff}})(\hat b ^\dagger -\hat b)\}}$, this contribution translates into a qubit tunneling term $\propto \hat \sigma _x$, giving rise to a sub-rotation of the effective frame. The resulting dynamics complies with the envelope defined by the ideal Hamiltonian with the tunneling term absent and therefore maps to the ideal quantum Rabi model, leaving its dynamics qualitatively unaffected. The transformations described are detailed in Supplementary Note 1, with master equation simulations supporting these statements in Supplementary Figure 6. In Fig.~\ref{fig:revival}(d) we made use of the topological symmetry of simulations with initial qubit states $\Ket{\mathrm{g}}$, $\Ket{\mathrm{e}}$, by subtracting two successive measurements with the qubit prepared in its eigenstates $\Ket{\mathrm{g}}$, $\Ket{\mathrm{e}}$, respectively, in order to cancel out the additional dispersive shift induced by the bosonic mode. As described in the Methods, we obtain the population evolution of the bosonic mode, depicted in Fig.~\ref{fig:revival}(e), by summing two successive measurements with the qubit prepared in $\Ket{\mathrm{g}}$, $\Ket{\mathrm{e}}$, respectively. We can infer its effective population by a fit to master equation simulations in the absence of a parasitic drive of the bosonic mode. Since the maximum population is around unity while the qubit is in the equatorial state, the non-conservation of the total excitation number is apparent.

While the phase of the qubit Bloch vector is not well defined for initial states $\Ket{\mathrm{g}}$, $\Ket{\mathrm{e}}$, the qubit state carries phase information when prepared on the equatorial plane of the Bloch sphere via a $\pi/2$ pulse. Figures~\ref{fig:revival}(f)-(i) show the qubit time evolution with varying relative phase $\varphi_1$ between initial state and applied drive, plotted in the original qubit basis, as calibrated in a Rabi oscillation experiment. Experimentally, the orientation of the coordinate system is set by the first microwave pulse and we apply the Rabi drive with a varying relative phase $\varphi_1$, corresponding to the angle between qubit Bloch vector and rotation axis of the drive in the equatorial plane. When both are perpendicular, $\varphi_1=\pm \pi/2$, similar oscillations including the state revival can be observed, assuming a steady state in the equatorial plane. For the case where $\varphi_1=0,\pi$, qubit oscillations in the laboratory frame are suppressed while the baseline is shifted up or down due to the detuning of the Rabi drive. The substructure emerges from the swap interaction term between qubit and bosonic mode that may be regarded as a perturbation as $\eta_1\gg g$. Classical master equation simulations confirm that the basis shift, dependent on the prepared initial qubit state, is enhanced by the presence of the second excited transmon level and by a spectral broadening of the applied Rabi drive. The experimentally observed shift is not entirely captured by the classical simulation which we attribute to missing terms in the master equation that may be related to qubit tuning pulses and are unknown at present. See Supplementary Note 4 for a further discussion of the effect. Dependent on $\varphi_1$, we observe a varying maximum photon population of the bosonic mode in classical simulations and indicated in the measured dispersive shift of the readout resonator. The qubit population as depicted in Fig.~\ref{fig:revival}(f)-(i) is retrieved from measured raw data by subtracting the contribution of the bosonic mode. A deviation of the effective qubit basis is likewise observed for preparing the qubit in one of its eigenstates $\Ket{\mathrm{g}}$, $\Ket{\mathrm{e}}$.

\subsection*{Full quantum Rabi model}
\begin{figure} 
\includegraphics{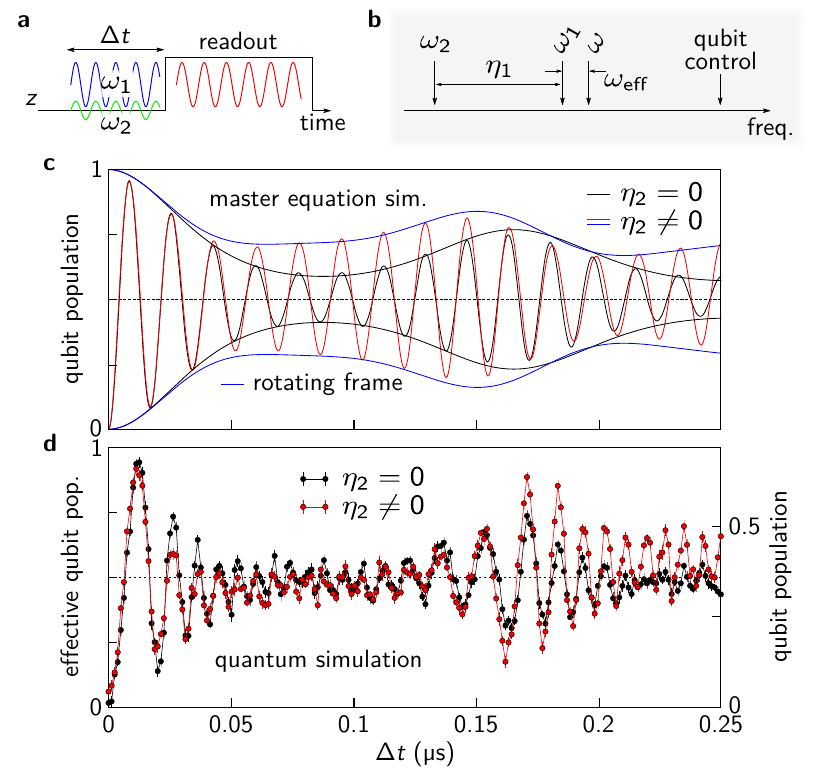}
\caption{\textbf{Simulation of the full quantum Rabi model} (a) Schematic pulse sequence used in the experiment. (b) Overview on the relative frequencies of the bosonic mode and the applied drives. The constraint $\eta_1=\omega_1-\omega_2$ is sketched. (c) Master equation simulations for vanishing qubit term $\eta_2=0$ (black) and with non-vanishing qubit term $\eta_2>0$ (red). The blue line corresponds to the classical simulation for $\eta_2/2\pi =\SI{3}{MHz}$. (d) Quantum simulation for equal parameters. The dispersive shift of the readout resonator induced by the bosonic mode is subtracted based on classically simulated data. Error bars denote a statistical standard deviation as detailed in the Methods.}
\label{fig:eta2}
\end{figure}

In order to simulate the full quantum Rabi model including a non-vanishing qubit energy term we switch on the second drive, $\eta_2\neq 0$, see Fig.~\ref{fig:eta2}(a). Quantum simulations are performed with the qubit initially in $\Ket{\mathrm{g}}$, subject to thermal excess population. The drive tones are up-converted in two separate IQ mixers while sharing a common local oscillator input to preserve their relative phase relation. For the simulation scheme to be valid, we need to fulfill the constraint $\omega_2=\omega_1-\eta_1$, see the schematics in Fig.~\ref{fig:eta2}(b). This is achieved by initially applying a simulation sequence with $\eta_2=0$ in order to obtain the frequency equivalent of the Rabi frequency $\eta_1$ from a Fourier transformation of the qubit time evolution. Subsequently, we apply the same sequence with a finite $\eta_2$, $\varphi_1=\varphi_2$ and $\omega_2$ set by obeying the above constraint. Figure~\ref{fig:eta2}(c) shows a master equation simulation of the complete quantum Rabi model for $\eta_2=0$ (black) and $\eta_2\neq 0$ (red), respectively. The main difference is an emerging substructure between quantum revivals and an increase of the revival amplitude in the presence of the qubit energy term. The substructure before the first revival is not reproduced in measured data, see Fig.~\ref{fig:eta2}(d), which we attribute to ring up dynamics of the applied drives, such that the frequency constraint of the simulation scheme is not satisfied at small $\Delta t$. In addition, the parasitic drive of the bosonic mode contributes in parts to the suppression of the substructure. Convergence of the experimental simulation however can be observed better at later simulation times, where we observe an increase in the revival amplitude and more pronounced oscillations after the first revival, in agreement with the classical simulation. These signatures vanish in check measurements for intentionally violating the above constraint or applying the weak Rabi drive with a phase delay $\varphi_1\neq\varphi_2$, see Supplementary Note 7. We estimate the frequency equivalent of $\eta_2/2\pi\sim \SI{3}{MHz}$ via comparing the relative peak heights of both drive tones with a spectrum analyzer. With $\omega_{\mathrm{eff}}/2\pi =\SI{6}{MHz}$ we approach a regime where $2g_{\mathrm{eff}}/\sqrt{\omega_{\mathrm{eff}}\eta_2/2}>1$, marking the quantum critical point in the related Dicke model \cite{Nataf2010}.

The limitations imposed by the low coherence in the slowed down effective frame can be mitigated in a future experiment by employing a high-quality 3D cavity featuring a dc bias and a dedicated Rabi drive antenna coupling to the qubit. Fast tuning pulses may be realized by making use of the ac Stark shift induced by an off-resonant tone. A device with stronger suppression of parasitic couplings to the bosonic mode would not further require a classical post processing, which allows to extend the presented scheme to regimes where classical simulations become very inefficient.

\section*{Discussion}
We have demonstrated analog quantum simulation of the full quantum Rabi model in the ultra-strong and close deep strong coupling regime. The distinct quantum state collapse and revival signature in the qubit dynamics was observed, validating the experimental feasibility of the proposed scheme \cite{Ballester2012}. The main limitation of the scheme is an effective slowing down of the system dynamics, while the laboratory frame dissipation rates are maintained in the synthesized frame. In analogy to the measure of cooperativity in standard QED, we find the ratio $g_{\mathrm{eff}}/\sqrt{\kappa/T_1}\sim 30$, rendering the qubit and bosonic mode decay rates an ultimate limitation for the simulation quality. The decelerated system dynamics in the effective frame however allows for the observation of quantum revivals on a timescale of $\sim\SI{100}{ns}$, while the revival rate in the laboratory frame USC quantum Rabi model is on an sub-nanosecond scale, being experimentally hard to resolve. The small transmon anharmonicity limits the Rabi frequency to below $\sim\SI{100}{MHz}\sim 0.3|\alpha|$ in order to avoid higher level populations and suppress parasitic coupling to the bosonic mode. The accessible coupling regime is not limited by the simulation scheme, however we can experimentally observe quantum revivals only up to a coupling regime where $g_{\mathrm{eff}}/\omega_{\mathrm{eff}}\sim 0.6$ due to the finite coherence in our circuit. 

While the presented dynamics can still be efficiently simulated on a classical computer, a true quantum supremacy will onset when incorporating more harmonic modes, leading to an exponential growth of the joint Hilbert space. Substituting the single quantized mode by a continuous bosonic bath renders our setup a viable tool for investigating the spin boson model in various coupling regimes, which recently attracted experimental interest in the context of quantum simulations \cite{Haeberlein2015,Forn-Diaz2016}. The presented simulation scheme can be applied for a continuum of modes, such that an engineered bath in a restricted frequency band is collectively shifted by the applied Rabi drive frequency. This can become a route to address the infrared cutoff issue in a tailored bosonic bath and to observe a quantum phase transition in the spin boson model.

\section*{Methods}

\subsection*{Experimental technique}
The quantum circuit is mounted in an aluminum box and cooled below $\sim\SI{50}{mK}$. It is enclosed in a cryoperm case for additional magnetic shielding. Qubit preparation and manipulation microwave pulses are generated by heterodyne single sideband mixing and applied to the same transmission line used for readout. In order to ensure phase control of the drive tones with respect to the qubit Bloch sphere coordinate system fixed by the first excitation pulse, we use a single microwave source for qubit excitation and the drives required by the simulation scheme. Different pulses are generated by heterodyne IQ mixing with separate IQ frequencies and amplitudes. The bosonic mode resonator is located far away from the transmission line which reduces parasitic driving. Readout of the qubit state is performed dispersively by means of a separate readout resonator located at $\omega_{\mathrm{r}}/2\pi=\SI{8.86}{GHz}$ in a projective measurement of the $\hat\sigma _z$ operator with a strong readout pulse of $\SI{400}{ns}$ duration. Further details on the experimental setup are given in Supplementary Note 3.

\subsection*{Protocol for extracting the qubit population}
In the simulation experiments presented in Fig.~\ref{fig:revival},~\ref{fig:eta2}, we note a modulated low-frequency bulge in the recorded dispersive readout resonator shift that does not agree with the expected qubit population evolution. By comparing with the classical master equation simulation, we can recognize the population evolution of the bosonic mode which reflects the governing fundamental frequency $\omega_{\mathrm{eff}}$ of the effective Hamiltonian. By simulating the full circuit Hamiltonian including qubit, bosonic mode and readout resonator, we find that the effect is induced by an additional photon exchange coupling $f$ between the bosonic mode and the readout resonator. The coupling is facilitated by the electric fields of the resonators and is potentially mediated by the qubit. See Supplementary Note 5 for the complete system Hamiltonian. In the diagonalized subspace of the two resonators, the bosonic mode can induce a cross-Kerr like photon number dependent shift $\propto f^2$ on the harmonic readout resonator as it inherits nonlinearity from the qubit. By adding or subtracting two subsequent simulation traces with the qubit prepared in either of the initial states $\Ket{\mathrm{g}}$, $\Ket{\mathrm{e}}$, we can isolate the signals corresponding to the population of the qubit and the bosonic mode. This measurement protocol is based on the symmetry of the qubit signal for preparing eigenstates, while the bosonic mode induced shift is always repulsive and does not change its sign. The photon exchange coupling $f$ therefore provides indirect access to the population of the bosonic mode without a dedicated readout device available. Specifically monitoring the population of the bosonic mode and performing a Wigner tomography would highlight another hallmark signature of the USC regime, namely the efficient generation of non-classical cavity states \cite{Langford2016}. Due to a lack of such a symmetry in case the relative phases of the Rabi drives are relevant, the qubit population can be retrieved from measured raw data based on the expectation for the bosonic mode population as obtained from the classical master equation simulation. In this procedure, the dispersive shift $\propto f^2$ remains as the only free fit parameter. See Supplementary Note 5 for more details on the described protocol.

\subsection*{Data acquisition}
We readout the qubit state by observing the dispersive shift of the readout resonator which is acquired via a $\SI{400}{ns}$ long readout pulse. Full time traces, recording the readout pulse, are $2\times 10^3$ fold pre-averaged per trace on our acquisition card. Successively, the data is sent to the measurement computer where we extract the IQ quadratures by Fourier transformation. We typically average over $\sim 30$ acquired traces to obtain a reasonable signal to noise ratio. Due to the reflection setup, most information is stored in the phase quadrature of the recorded signal. The given error bars represent the standard deviation of the mean, as calculated from the pre-averaged data points and propagated according to Gauss.

\subsection*{Data availability}
The data that support the findings of this study are available from the corresponding author upon reasonable request.

\begin{acknowledgments}
The authors acknowledge valuable discussions with G. Romero, M.-J. Hwang, I. Pop, U. Vool, J. Pedernales and E. Solano. We are grateful to L. Radtke and S. Diewald for support during sample fabrication and A. Lukashenko for assistance in cryostat operation. This work was supported by the European Research Council (ERC) within consolidator grant No. 648011 and through the KIT Nanostructure Service Laboratory (NSL). This work was also supported in part by the Ministry for Education and Science of the Russian Federation via NUST MISIS under contracts K2-2016-063 and K2-2016-051. J.B. acknowledges financial support by the Landesgraduiertenf\"orderung (LGF) of the federal state Baden-W\"urttemberg and by the Helmholtz International Research School for Teratronics (HIRST). A.Sch. acknowledges financial support by the Carl-Zeiss-Foundation.
\end{acknowledgments}

\section*{Author contributions}
J.B. designed and fabricated the device, with input from M.W. and H.R. J.B. performed the measurements with support by A.Sch. and A.S. J.B. carried out data analysis and numerical simulations with contributions from M.M. J.B. wrote the manuscript with input from all coauthors. M.W. and A.U. supervised the project.

\section*{Competing financial interests}
The authors declare that they have no competing financial interests.

\end{document}


\title{Analog quantum simulation of the Rabi model in the ultra-strong coupling regime\\Supplemental Material}

\author{Jochen Braum\"uller}
\email{jochen.braumueller@kit.edu.}
\affiliation{Physikalisches Institut, Karlsruhe Institute of Technology, 76131 Karlsruhe, Germany}

\author{Michael Marthaler}
\affiliation{Institut f\"ur Theoretische Festk\"orperphysik, Karlsruhe Institute of Technology, 76131 Karlsruhe, Germany}

\author{Andre Schneider}
\author{Alexander Stehli}
\author{Hannes Rotzinger}
\affiliation{Physikalisches Institut, Karlsruhe Institute of Technology, 76131 Karlsruhe, Germany}

\author{Martin Weides}
\affiliation{Physikalisches Institut, Karlsruhe Institute of Technology, 76131 Karlsruhe, Germany}
\affiliation{Physikalisches Institut, Johannes Gutenberg University Mainz, 55128 Mainz, Germany}

\author{Alexey V. Ustinov}
\affiliation{Physikalisches Institut, Karlsruhe Institute of Technology, 76131 Karlsruhe, Germany}
\affiliation{Russian Quantum Center, National University of Science and Technology MISIS, Moscow 119049, Russia}

\date{\today}

\section{\label{sec:theory}Theoretical basics of the simulation scheme}

The simulation scheme we detail here follows a proposal published in Ref.~\cite{Ballester2012}. The quantum Rabi Hamiltonian to be constructed reads
\begin{equation}
\frac{\hat H}{\hbar} =\frac{\epsilon} 2 \hat\sigma _z+\omega\hat b ^\dagger \hat b+g\hat\sigma _x\left(\hat b ^\dagger +\hat b\right),
\label{eq:HRabi}
\end{equation}
with $\epsilon$ the qubit energy splitting, $\omega$ the resonator frequency and $g$ the transversal coupling strength. $\hat\sigma _i$ are Pauli matrices with $\hat\sigma _z\Ket{\mathrm{g}}=\Ket{\mathrm{g}}$ and $\hat\sigma _z\Ket{\mathrm{e}}=-\Ket{\mathrm{e}}$, where $\Ket{\mathrm{g}}$, $\Ket{\mathrm{e}}$ are the qubit groundstate and first excited state, respectively. We neglect the qubit tunneling matrix element $\Delta$ such that the qubit Hamiltonian remains
\begin{equation}
\frac{\hat H _{\mathrm{q}}}{\hbar}=\frac{\epsilon} 2 \hat\sigma _z +\frac{\Delta} 2 \hat\sigma _x \approx \frac{\epsilon} 2 \hat\sigma _z.
\end{equation}
$\Delta \ll \epsilon$ is valid for the transmon qubit since hopping between wells of the Josephson potential is suppressed in the spirit of a WKB approximation \cite{Leggett1987}. $\hat b ^\dagger$ ($\hat b$) are creation (annihilation) operators in the resonator Fock space. Throughout this derivation, operators acting on either the qubit or the resonator subspace of the joint Hilbert space are written without the formally correct tensor product with the unity in the other subspace, such that $\mathbbm{1}_{2}\otimes\hat b ^\dagger \hat b \equiv \hat b ^\dagger \hat b$, $\hat\sigma _z\otimes\mathbbm{1}_{\mathrm{r}} \equiv \hat\sigma _z$. $\mathbbm{1}_2$ denotes unity in the qubit Hilbert space while $\mathbbm{1}_{\mathrm{r}}$ denotes unity in the bosonic mode Hilbert space of dimension $r$. Likewise, the tensor product symbol is omitted for clarity.

The qubit and the bosonic oscillator mode are physical elements of the quantum simulator implemented on chip. Since the geometric coupling $g$ on chip is small compared to the mode energies, $g/\epsilon\ll 1$ and $g/\omega\ll 1$, the rotating wave approximation (RWA) is valid and Supplementary Equation~\eqref{eq:HRabi} takes the form of the Jaynes-Cummings Hamiltonian \cite{Jaynes1963}. The construction of the effective Hamiltonian relies on the application of two transversal microwave tones acting on the qubit. The key feature is a renormalization of the mode energies $\epsilon$, $\omega$ by experimental parameters in the frame rotating with the first dominant drive. Even though a RWA may be used to simplify the ordinary Jaynes-Cummings Hamiltonian as implemented on chip, the RWA breaks down for the quantum Rabi Hamiltonian in the ultra-strong coupling (USC) regime and beyond. Therefore the complete coupling term $g\hat\sigma _x(\hat b ^\dagger +\hat b)$ needs to be preserved. The Jaynes-Cummings Hamiltonian in the laboratory frame with the two drives applied takes the form
\begin{equation}
\frac{\hat H}{\hbar} =\frac{\epsilon} 2 \hat\sigma _z+\omega\hat b ^\dagger \hat b +g\left(\hat\sigma _-\hat b ^\dagger +\hat\sigma _+\hat b\right)\nonumber +\hat\sigma _x\eta_1\cos(\omega_1 t+\varphi_1)+\hat\sigma _x \eta_2\cos(\omega_2 t+\varphi_2),
\label{eq:HDrives}
\end{equation}
with $\eta_i$ the amplitudes, $\omega_i$ the frequencies and $\varphi_i$ the relative phase of drive $i$. In the following derivation we set $\varphi_i=0$ without loss of generality. We use Pauli's ladder operators $\hat\sigma_{\pm} =1/2\left(\hat\sigma _x\pm \I\hat\sigma _y\right)$. The unitary transformation $\hat U$ for changing into the frame rotating with the dominant drive $\eta_1$ reads
\begin{equation}
\hat U = \exp\left\{\I\omega_{1}t\left(\hat b ^\dagger \hat b+\frac{1}{2}\hat\sigma _z\right)\right\}.
\label{eq:U}
\end{equation}
Performing the transformation $|\tilde{\psi}\rangle=U|\psi\rangle$ of eigenstates $|\psi \rangle$ leads to a transformed Hamiltonian $\tilde H$ according to
\begin{equation}
\tilde{H}=\hat U\hat H\hat U^{\dagger}-\I\hat U\dot{\hat U}^{\dagger}.
\label{eq:Ht}
\end{equation}
The transformation of Supplementary Equation~\eqref{eq:HDrives} yields
\begin{widetext}
\begin{align}
\frac{\tilde H}{\hbar} = & \frac{\epsilon}{2}\hat\sigma _z +\omega\hat b ^\dagger \hat b +g\left(\hat\sigma _+ e^{\I\omega_1 t} +\hat\sigma _- ^{-\I\omega_1 t}\right) \left(e^{\I\omega_{1}t}\hat b ^\dagger+e^{-\I\omega_{1}t}\hat b\right)\nonumber \\
& +\eta_1\left(\hat\sigma _+ e^{\I\omega_1 t} +\hat\sigma _- e^{-\I\omega_1 t}\right)\cos\omega_1t +\eta_2\left(\hat\sigma _+ e^{\I\omega_1 t} +\hat\sigma _- e^{-\I\omega_1 t}\right)\cos\omega_2t -\omega_{1}\left(\hat b ^\dagger \hat b+\frac{1}{2}\hat\sigma _z\right) \\
= & \frac{1}{2}\left(\epsilon-\omega_{1}\right)\hat\sigma _z +\left(\omega-\omega_{1}\right)\hat b ^\dagger \hat b +g\left(\hat{\sigma}_{-}\hat b ^\dagger+\hat{\sigma}_{+}\hat b\right) +\frac{\eta_1}{2}\left(\hat{\sigma}_{+}+\hat{\sigma}_{-}\right) +\frac{\eta_2}2\left(\hat\sigma _+ e^{\I(\omega_1-\omega_2)t}+\hat\sigma _- e^{-\I(\omega_1-\omega_2)t}\right)\nonumber \\
  & +g\left(\hat\sigma _+e^{2\I\omega_{1}t}\hat b ^\dagger+\hat\sigma _-e^{-2\I\omega_{1}t}\hat b\right) +\frac{\eta_1}{2}\left(\hat\sigma _+e^{2\I\omega_{1}t}+\hat\sigma _-e^{-2\I\omega_{1}t}\right) +\frac{\eta_2}{2}\left(\hat{\sigma}_{+} e^{\I(\omega_1+\omega_2)t} +\hat{\sigma}_{-} e^{\I(-\omega_1-\omega_2)t}\right).
\label{eq:Htilde}
\end{align}
\end{widetext}
Terms of the form $e^{X}Ye^{-X}$ are calculated using the power series expansion of the exponential function, also called Hadamard lemma. For $X=\I\omega_1t\hat b ^\dagger\hat b ,\: Y=\hat b ^\dagger$, 
\begin{align}
e^{\I\omega_1t\hat b ^\dagger\hat b}\hat b ^\dagger e^{-\I\omega_1t\hat b^\dagger\hat b} &=e^{\I\omega_1t}\hat b ^\dagger\\
e^{\I\omega_1t\hat b ^\dagger\hat b}\hat b e^{-\I\omega_1t\hat b^\dagger\hat b} &=e^{-\I\omega_1t}\hat b
\label{eq:hadlemma}
\end{align}
since
\begin{equation}
\sum_{m=0}^\infty \frac 1 {m!} \left[\I\omega_1 t \hat b ^\dagger\hat b ,\hat b ^\dagger \right]_m=\hat b ^\dagger \sum_{m=0}^\infty \frac 1 {m!} \left(\I\omega_1 t\right)^m=\hat b ^\dagger e^{\I\omega_1 t}.
\end{equation}

$e^{\frac \I 2 \omega_1 t\hat\sigma _z}$ and $\hat\sigma_z$ clearly commute, while
\begin{equation}
e^{\frac \I 2 \omega_1 t\hat\sigma _z} \hat\sigma _x e^{-\frac \I 2 \omega_1 t\hat\sigma _z} = \hat\sigma _{+} e^{\I\omega_1 t} + \hat{\sigma}_{-}e^{-\I\omega_1 t}
\end{equation}
using $e^{\text{diag}(a,b)} = \text{diag}(e^a,e^b)$. Terms in the last line in Supplementary Equation~\eqref{eq:Htilde} are omitted in the following within the RWA, valid for $\eta_1/2\omega_1\ll 1$. This is a good approximation as $\eta_1$ is bound in the experiment by the qubit anharmonicity $|\alpha| \lesssim \SI{350}{MHz}$.

The $\eta_1$ term in the transformed Hamiltonian Supplementary Equation~\eqref{eq:Htilde} is the most significant term, which justifies to move to its interaction picture. Under the constraint $\eta_1\equiv\omega_1-\omega_2$, the Hamiltonian in the interaction picture becomes
\begin{widetext}
\begin{align}
e^{\I\frac{\eta_1}2 \hat\sigma _x t} \left[\frac{\tilde H}{\hbar}-\frac{\eta_1}2 \hat\sigma _x\right]e^{-\I\frac{\eta_1}2 \hat\sigma _x t} 
= &\frac{\eta_2}2 \left(\begin{array}{cc} \sin^2\eta_1t & \cos\eta_1t+\I\sin2\eta_1t\\ \cos\eta_1t-\I\sin2\eta_1t & -\sin^2\eta_1t \end{array}\right)\nonumber\\&
+\frac{1}{2}\left(\epsilon-\omega_{1}\right)\left(\begin{array}{cc} \cos\eta_1t & -\I\sin\eta_1t\\ \I\sin\eta_1t & -\cos\eta_1t \end{array}\right)+\left(\omega-\omega_1\right)\hat b ^\dagger\hat b\nonumber\\&
+g\left[\left(\begin{array}{cc}-\frac 1 2 \I\sin\eta_1t & \frac 1 2 (1+\cos\eta_1t)\\ \frac 1 2 (1-\cos\eta_1t) & \frac 1 2 \I\sin\eta_1t\end{array}\right) \hat b ^\dagger+\left(\begin{array}{cc}\frac 1 2 \I\sin\eta_1t & \frac 1 2 (1-\cos\eta_1t)\\ \frac 1 2 (1+\cos\eta_1t) & -\frac 1 2 \I\sin\eta_1t\end{array}\right) \hat b\right].
\label{eq:interactionpic}
\end{align}
\end{widetext}
Performing a time averaging in the spirit of a RWA casts Supplementary Equation~\eqref{eq:interactionpic} into the desired form of the quantum Rabi Hamiltonian
\begin{equation}
\frac{\hat H _{\mathrm{eff}}}{\hbar} = \frac{\eta_2}2\frac{\hat\sigma _z}2 +\omega_{\mathrm{eff}}\hat b ^\dagger \hat b +\frac g 2\sigma _x\left(\hat b ^\dagger +\hat b\right),
\label{eq:Hrot1eff}
\end{equation}
with $\omega_{\mathrm{eff}}\equiv \omega-\omega_1$ and noting that $\eta_1\gg \eta_2$.

In the experiment, the applied Rabi drives parasitically couple to the bosonic mode since the elements are degenerate during the simulation experiment and spatially close by in the circuit. This effect is accounted for by an additional term
\begin{equation}
\eta_{\mathrm{r}} \left(\hat b ^\dagger +\hat b\right)\cos(\omega_1 t)
\label{eq:parasitic}
\end{equation}
in the laboratory frame Hamiltonian, Supplementary Equation~\eqref{eq:HDrives}. $\eta_{\mathrm{r}}$ denotes the effective amplitude of the parasitic drive and we take into account the effect of only the dominant Rabi drive frequency $\omega_1$. Performing the transformation $\hat U$ yields
\begin{equation}
\frac{\eta_{\mathrm{r}}}2 \left(\hat b ^\dagger + \hat b\right) +\frac{\eta_{\mathrm{r}}}2\left(e^{2\I\omega_1 t}\hat b ^\dagger +e^{-2\I\omega_1 t}\hat b\right).
\label{eq:par2}
\end{equation}
Terms rotating with $e^{\pm 2\I\omega_1 t}$ ore omitted, resulting in a time independent drive term that is added to the Hamiltonian \eqref{eq:Hrot1eff} in the rotating frame. The effective Hamiltonian including the parasitic drive term reads
\begin{equation}
\frac{\hat H _{\mathrm{eff,p}}}{\hbar} = \frac{\eta_2}2\frac{\hat\sigma _z}2 +\omega_{\mathrm{eff}}\hat b ^\dagger \hat b +\frac g 2\sigma _x\left(\hat b ^\dagger +\hat b\right) +\frac{\eta_{\mathrm{r}}}2 \left(\hat b ^\dagger + \hat b\right).
\label{eq:Hrot2}
\end{equation}
By applying the unitary displacement transformation
\begin{equation}
\hat D =\exp{\left\{-\frac{\eta_{\mathrm{r}}}{2\omega_{\mathrm{eff}}}\left(\hat b ^\dagger -\hat b\right)\right\}}
\label{eq:displ}
\end{equation}
we can cast the Hamiltonian \eqref{eq:Hrot2} in the original form of the quantum Rabi Hamiltonian, including a qubit tunneling term $\propto \hat \sigma _x$,
\begin{align}
\hat D ^\dagger \frac{\hat H}{\hbar} \hat D &= \frac{\eta_2}2\frac{\hat\sigma _z}2 +\omega_{\mathrm{eff}}\left(\hat b ^\dagger -\frac{\eta_{\mathrm{r}}}{2\omega_{\mathrm{eff}}}\right)\left(\hat b -\frac{\eta_{\mathrm{r}}}{2\omega_{\mathrm{eff}}}\right) +\frac g 2\sigma _x\left(\hat b ^\dagger +\hat b -\frac{\eta_{\mathrm{r}}}{\omega_{\mathrm{eff}}}\right) +\frac{\eta_{\mathrm{r}}}2 \left(\hat b ^\dagger +\hat b -\frac{\eta_{\mathrm{r}}}{2\omega_{\mathrm{eff}}}\right)\\
&= \frac{\eta_2}2\frac{\hat\sigma _z}2 -g\frac{\eta_{\mathrm{r}}}{2\omega_{\mathrm{eff}}}\hat \sigma _x +\omega_{\mathrm{eff}}\hat b ^\dagger \hat b +\frac g 2\sigma _x\left(\hat b ^\dagger +\hat b\right) +\mathrm{const.},
\label{eq:displHeff}
\end{align}
using
\begin{equation}
\hat D ^\dagger \hat b ^\dagger \hat b \hat D = \hat D ^\dagger \hat b ^\dagger \hat D \hat D ^\dagger \hat b \hat D = \left(\hat b ^\dagger -\frac{\eta_{\mathrm{r}}}{2\omega_{\mathrm{eff}}}\right)\left(\hat b -\frac{\eta_{\mathrm{r}}}{2\omega_{\mathrm{eff}}}\right).
\end{equation}

\section{\label{sec:fab}Sample fabrication}

Sample fabrication was carried out in one single electron beam lithography step. The Josephson junctions are formed by shadow angle evaporation and a Dolan bridge technique with electrode film thicknesses of $\SI{30}{nm}$ and $\SI{50}{nm}$, respectively, resulting in a Al film thickness of $\SI{80}{nm}$ across the entire chip. The area of the Josephson junctions is designed to be $\SI{100}{nm}\times\SI{220}{nm}$, resulting in a critical current $I_{\mathrm{c}}=\SI{45}{nA}$ for a single junction. Al is evaporated at a chamber background pressure of about $\SI{3e-8}{mbar}$. We applied an Al metalization on the backside of the double-side polished intrinsic Si substrate as a ground reference for the microstrip elements on chip. Electric field coupling in the substrate due to the backside metalization accounts for roughly half of the qubit capacitance \cite{Braumueller2016}.

\section{\label{sec:expsetup}Experimental setup}

\subsection{\label{sec:setup}Microwave setup}
\begin{figure}[h] 
\includegraphics{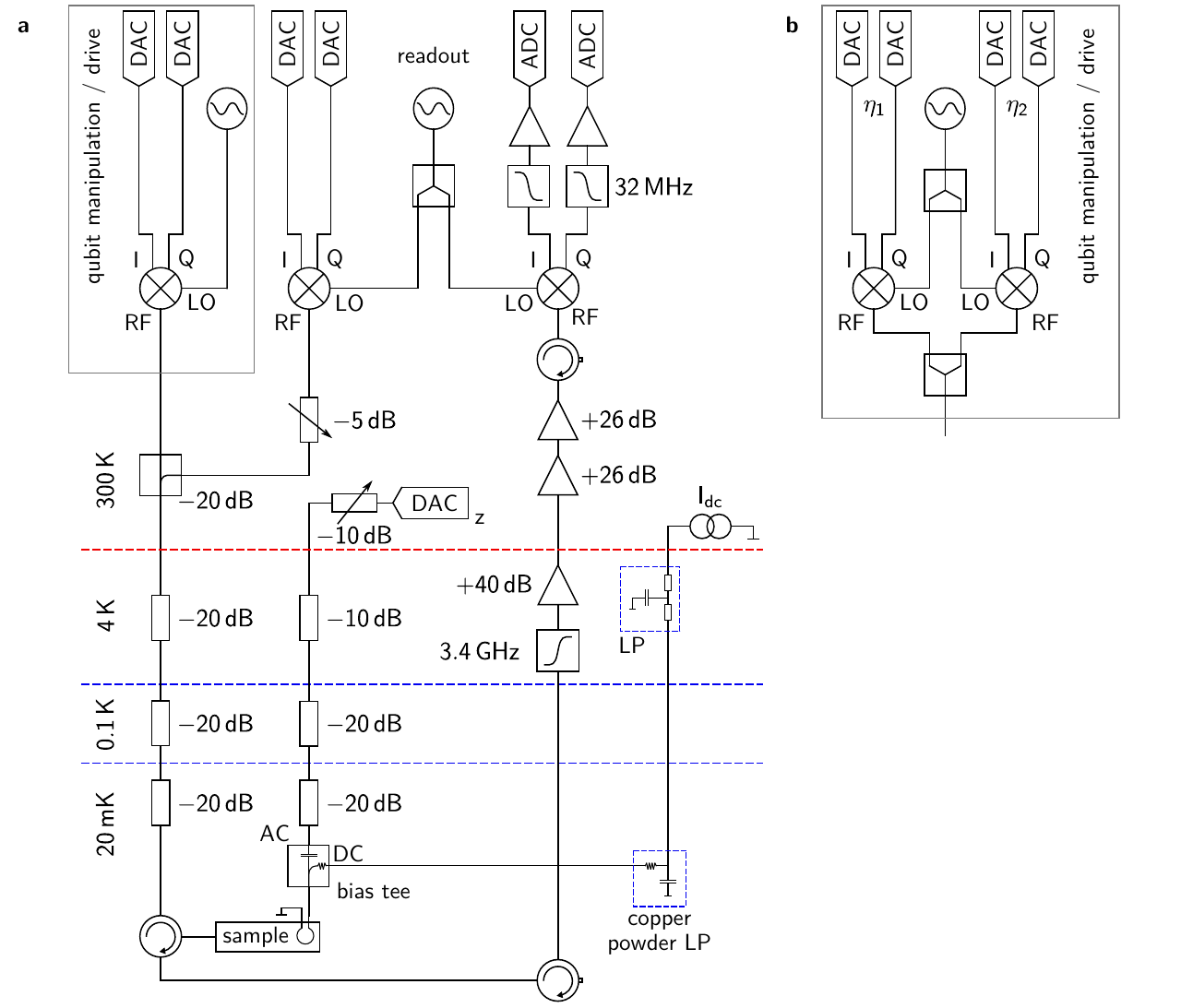}
\caption{\textbf{Schematic microwave setup used for the quantum simulation experiment} (a) Setup used for the experiments requiring only one Rabi drive tone. (b) Modification of the microwave setup (substituting the gray rectangle) for the application of two drive pulses with amplitudes $\eta_1$, $\eta_2$. The drive pulses are generated with two separate IQ mixers, sharing a common local oscillator (LO) input, and combined subsequently.}
\label{fig:setup}
\end{figure}

The schematic diagram of the measurement and microwave setup is depicted in \ref{fig:setup}. We readout the qubit state by observing a dispersive shift of the readout resonator which is acquired via a $\SI{400}{ns}$ long readout pulse. The resonator shift is extracted from the microwave reflection signal at a single-ended $50\,\mathrm{\Omega}$ matched transmission line that capacitively couples to the readout resonator. We frequency-convert the readout pulse by heterodyne single sideband mixing to eliminate parasitic population of the readout device during pulse-off time.

Qubit excitation and Rabi driving are performed by heterodyne mixing with respective IQ frequencies for different drive and excitation frequencies, using one single microwave source and one IQ mixer, see \ref{fig:setup}(a). This allows for the required phase control on the phases $\varphi_1$, $\varphi_2$ of the Rabi drives. In particular, we fix the idling time between initial excitation pulse and the onset of the Rabi drive, such that the acquired phase during that time is constant, and apply the Rabi drives with a constant relative phase $\varphi_i$ with respect to the phase used for the excitation pulse. We chose an LO frequency located $\SI{20}{MHz}$ or $\SI{65}{MHz}$ above the qubit control frequency, located at about $\omega/2\pi+\SI{95}{MHz}$. In the experiment with the second Rabi drive added, we generate the drives initially in separate IQ mixers that share a common LO input. We suppress phase errors by employing identical coaxial cables for the high-frequency lines prior to the combination of the drive pulses. The microwave pulses for $\hat X\hat Y$ control of the qubit are applied via the same transmission line used for readout.

The qubit transition frequency is adjusted by a dc current applied to the on-chip flux coil. High frequency noise is filtered at the $4\,\mathrm{K}$ stage with RCR type $\pi$-filters at about $25\,\mathrm{kHz}$ and on the base plate via an RC-element enclosed in copper powder \cite{Lukashenko2008}. Fast flux pulses for fast $\hat Z$ pulsing of the qubit are sent through a separate microwave line and combined with the offset current by means of a bias tee located at the base plate.

\subsection{\label{sec:biastee}Compensation for finite bias tee time constant}
\begin{figure}[h] 
\includegraphics{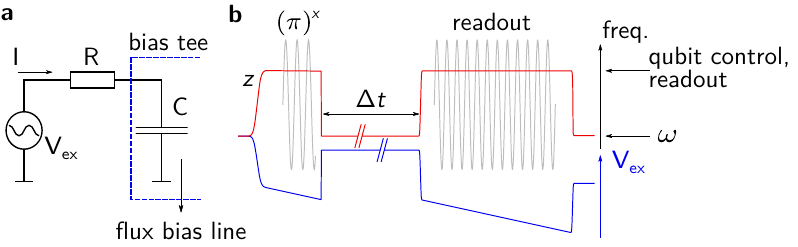}
\caption{\textbf{Compensation for finite time constant in the bias tee} (a) Simplified schematic circuit diagram prevailing the flux pulse generation. Microwave pulses of amplitude $V_{\mathrm{ex}}$ pass a resistor $R$ and charge the capacitor which is part of the bias tee. (b) Due to the finite time constant $\tau$ of the bias tee, voltage pulses of amplitude $V_{\mathrm{ex}}$ following the blue line are applied to its input such that the resulting flux through the on chip flux coil follows the ideal pulse sequence depicted in red.}
\label{fig:biastee}
\end{figure}

In order to combine dc and ac flux signals that are applied to the flux bias line without breaking the $\SI{50}{\Omega}$ impedance matching, we make use of a bias tee in the experiment. Due to a finite time constant $\tau$ of the bias tee, a continuous compensation for decharging effects is required in order to produce the desired pulse sequence in the flux bias line on chip. \ref{fig:biastee}(a) shows the schematic circuit diagram present during flux pulse generation. $V_{\mathrm{ex}}$ denotes the amplitude of the voltage pulse, $R$ is the line resistance in front of the bias tee and $C$ is the relevant capacitance of the bias tee. From Kirchhoff's law we can write
\begin{equation}
RI+\frac Q C = V_{\mathrm{ex}},
\end{equation}
with $I$ the current that is admitted to the flux bias line. The condition of constant current follows from requesting
\begin{equation}
\frac{\rm{d}I}{\rm{d}t} =\frac 1 R \frac{\rm{d}}{\rm{d}t}V_{\mathrm{ex}} -\frac 1 {RC}I \stackrel{!}{=} 0,
\end{equation}
which yields
\begin{equation}
\frac{\rm{d}}{\rm{d}t}V_{\mathrm{ex}} = \frac I C.
\label{eq:dVex}
\end{equation}
Since $I$ is not a function of time according to the initial claim, Supplementary Equation~\eqref{eq:dVex} can be integrated yielding
\begin{equation}
V_{\mathrm{ex}}=\frac 1 C It +\mathrm{const.}\,\propto t.
\end{equation}
This shows that the required correction of the externally applied voltage $V_{\mathrm{ex}}$ is linear in time $t$ with a slope proportional to $1/\tau$. See the pulse sequence applied (blue) and the resulting pattern (red) in \ref{fig:biastee}(b). During pulse-off time, $V_{\mathrm{ex}}$ has to be kept constant and no further correction is required. The described compensation cannot be performed for longer than approximately $\SI{1}{\micro s}$ within one continuous pulse sequence, since the output stage of our pulse generator eventually saturates. With additional amplification of the generated signal, the current is increasing linearly during compensation time, implying an experimental limitation as higher currents can ultimately heat the cryostat. The utilized pulse sequence therefore scales nicely for perspective longer simulation times $\Delta t$ since a compensation during $\Delta t$ is not required.

We calibrated the time constant of the bias tee used in the experiment to be $\tau =\SI{0.7}{\micro s}$.

\subsection{\label{sec:table}Summary of the device parameters}

\begingroup
\squeezetable
\begin{table}[h] 
\begin{center}
\begin{ruledtabular}
\begin{tabular}{cccccccccccc}
$\omega/2\pi$ & $E_{\mathrm{J}}/E_{\mathrm{C}}$ & $\alpha/h$ & $1/T_1$ & $\kappa$ & $\omega_{\mathrm{r}}/2\pi$ & $\omega_{\mathrm{eff}}/2\pi$ & $\eta_1/2\pi$ & $\eta_2/2\pi$ \\
\hline\\
$\SI{5.948}{GHz}$ & $50$ & $\SI{-0.36}{GHz}$ & $(0.2\pm 0.12)\times 10^ 6\,\mathrm{s^{-1}}$ & $(3.9\pm 0.14)\times 10^ 6\,\mathrm{s^{-1}}$ & $\SI{8.86}{GHz}$ & $<\SI{8}{MHz}$ & $\sim\SI{50}{MHz}$ & $\sim\SI{3}{MHz}$\\
\end{tabular}
\caption{\textbf{List of relevant device and simulation parameters.} Errors are extracted from the fit of the vacuum Rabi oscillations in Fig.~2 in the main text.}
\label{tab:paramsoverview}
\end{ruledtabular}
\end{center}
\end{table}

\begin{table}
\begin{center}
{\renewcommand{\arraystretch}{2}   
\begin{tabular}{@{\hspace{10pt}} c @{\qquad} c @{\qquad} c @{\hspace{10pt}}}
\toprule
spectroscopy & vacuum Rabi & master eq. simulation\\
\colrule
$\SI{3.9}{MHz}$ & $\SI{4.3}{MHz}$ & $\SI{5.5}{MHz}$\\
\botrule
\end{tabular}
}
\caption{\textbf{Geometric coupling $g/2\pi$ between qubit and bosonic mode.} Values are obtained from the spectroscopic measurement in \ref{sec:spec}, the vacuum Rabi experiment in Fig.~2 in the main text and the value used in master equation simulations.}
\label{tab:ggeo}
\end{center}
\end{table}
\endgroup

\ref{tab:paramsoverview} lists the relevant parameters of the simulation device that were found experimentally and used in the master equation simulations, together with drive parameters typically used for the quantum simulation.

Values for the geometric coupling strength $g$ are summarized in \ref{tab:ggeo}. Values from the spectroscopy experiment and the vacuum Rabi experiment are in reasonable agreement. The geometric coupling appears to be effectively enhanced during the simulation experiment, which could be due to a parameter drift during successive cool-downs. In order to achieve a better agreement between experiment and master equation simulation, we use a slightly increased value for the geometric coupling strength in classical simulations. We believe that the effective modification of this parameter is a consequence of the discussed parasitic driving of the bosonic mode in the experiment. We checked that this has no qualitative influence on the results, in particular the non-conservation of the total excitation number remains evident.

\section{\label{sec:calib}Calibration of the qubit basis and baseline shift}

\subsection{\label{sec:Rabi}Calibration of the qubit basis}

\begin{figure} 
\pagebreak
\includegraphics{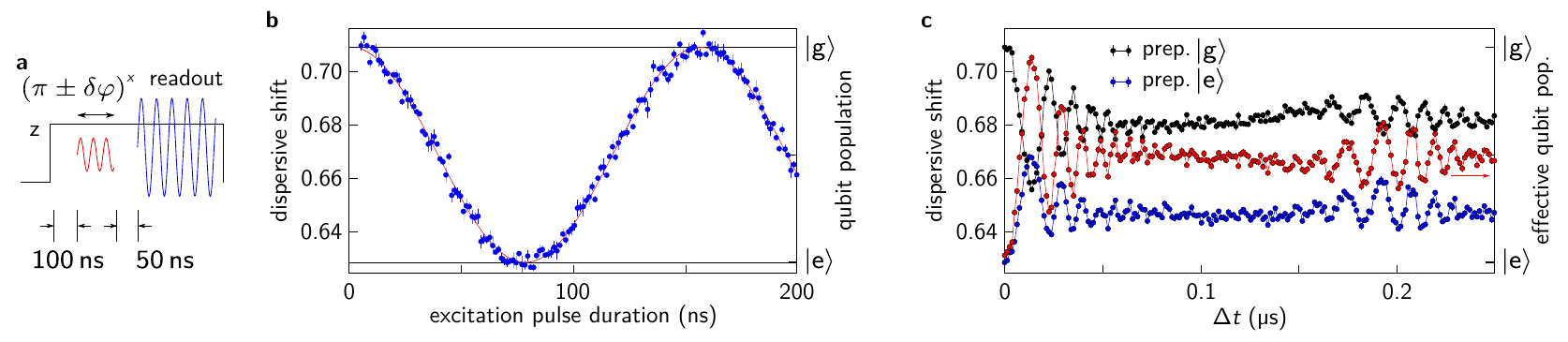}
\caption{\textbf{Calibration of the qubit basis} (a) Schematic pulse sequence used for the calibration. Qubit and bosonic mode are on resonance initially. A fast flux pulse is applied to tune the qubit out of resonance for excitation and readout. The length of the excitation pulse corresponding to the rotation angle $\varphi$ around the Bloch sphere is varied during the calibration. (b) Qubit Rabi oscillations using the $z$ pulsing sequence used in the simulation experiment and with a sinusoidal fit (red) applied. We identify the computational qubit basis spanned by the groundstate $\Ket{\mathrm{g}}$ and the first excited state $\Ket{\mathrm{e}}$. (c) Dispersive shift of the readout resonator for a simulation sequence with $\omega_{\mathrm{eff}}/2\pi=\SI{5}{MHz}$ and the qubit prepared in $\Ket{\mathrm{e}}$ (blue) or $\Ket{\mathrm{g}}$ (black). The faintly visible bulge cancels by evaluating their difference, plotted in red. Error bars throughout the figure denote a statistical standard deviation as detailed in the Methods of the main text.}
\label{fig:rabi}
\end{figure}

Prior to the experiments presented in the main text, we calibrated the qubit basis spanned by eigenstates $\Ket{\mathrm{g}}$, $\Ket{\mathrm{e}}$. This is of relevance for the data analysis detailed in \ref{sec:qretrieval} and to demonstrate the base line shifts observed in Fig.~3 in the main text. The calibration is performed by Rabi spectroscopy including the fast $z$ pulsing scheme as adopted in successive quantum simulation experiments. The applied pulse sequence and measured Rabi oscillations are depicted in \ref{fig:rabi}. The length of the applied excitation pulse is given on the horizontal axis in \ref{fig:rabi}(b). Fundamental qubit states (black lines) can be mapped to dispersive shifts of the readout resonator with the bosonic mode in its groundstate and off-resonant.

\subsection{\label{sec:basisshift}Qubit basis shift}

The qubit state is expected to assume an incoherent steady state on the equator of the Bloch sphere for long simulation times $\Delta t$ due to the finite coherence in our circuit. However, we notice a shift in the steady state qubit population dependent on the initially prepared qubit state. This is apparent in \ref{fig:rabi}(c), showing measurements for the qubit prepared in $\Ket{\mathrm{e}}$ (blue) and $\Ket{\mathrm{g}}$ (green) and in Fig.~3(f)-(i) in the main text. We attribute this effect to a change in the effective qubit basis, which is not captured by the master equation simulations performed. We additionally conjecture that the basis shifts are caused by an effective tilt of the qubit Bloch sphere as an artifact of the frequency tuning in experiment prior to applying the Rabi drives. We isolate the effect as an initialization issue since the basis shift cancels out in good approximation when averaging the mutually antiparallel simulation sequences in Fig.~3(f)-(i).

\section{\label{sec:simulation}Classical master equation simulation}

Numerical simulations are based on a master equation solver provided by the QuTiP package \cite{Johansson2012,Johansson2013} for python. The time evolution of a given initial state or density matrix is calculated by solving the von Neumann equation associated with the given system Hamiltonian in the absence of dissipation. For including losses to an imperfect environment of the system Hamiltonian, the time evolution of the density matrix is calculated via the Lindblad master equation.

Classical simulations in the main text include a finite lifetime of the qubit of about $\SI{5}{\micro s}$, a dephasing time of about $\SI{0.5}{\micro s}$ and a bosonic mode inverse lifetime of $\kappa=3.9\times 10^ 6\,\mathrm{s^{-1}}$. The Fock space of the bosonic mode is truncated at a photon number of $\SI{25}{}$, since higher excitation numbers were found to not play a significant role. The transmon qubit is treated as a three-level system with the experimentally found anharmonicity $\alpha/2\pi=\SI{-350}{MHz}$. We use a transversal qubit coupling operator
\begin{equation}
\sum_{ij}\hat g_{ij}^x\equiv \sum_{ij}\frac{g_{ij}}{g_{01}}\Ket{i}\Bra{j},
\end{equation}
with the coupling matrix elements $g_{ij}$ found by evaluating the Cooper pair number operator in the charge basis \cite{Koch2007,Braumuller2015}.

\subsection{\label{sec:simscheme}Verification of the simulation scheme}

\begin{figure}[H] 
\begin{centering}
\includegraphics{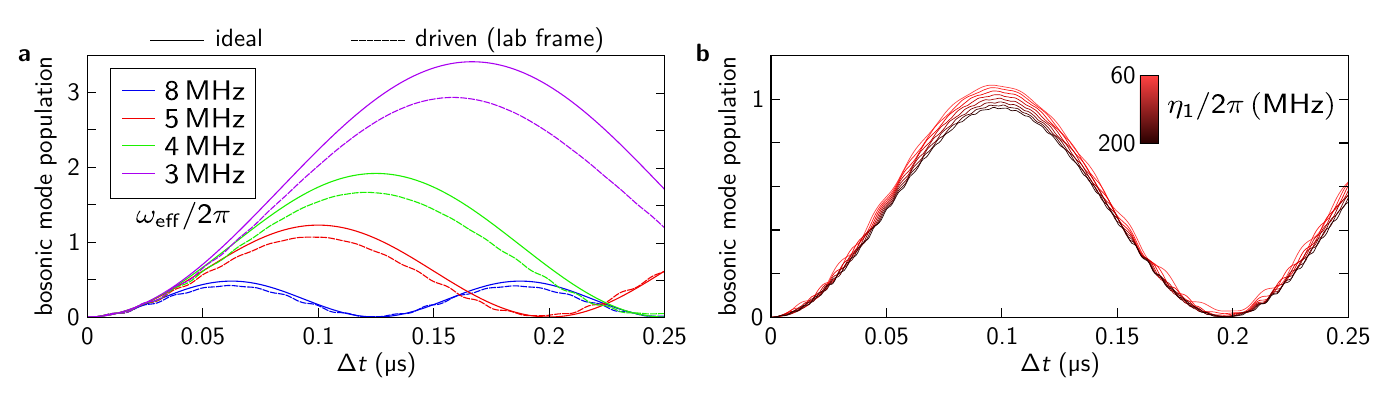}
\caption{\textbf{Verification of the simulation scheme for the ideal system} (a) Time evolution of the bosonic mode population for a simulation sequence with $\eta_2=0$. We compare its population in the ideal quantum Rabi Hamiltonian (solid line) with the population of the bosonic mode in the laboratory frame with the drive applied (dashed line). Despite the fact that an infinite energy reservoir is supplied by the drive, the population follows the expected one rather well. This remains also true for varying the drive amplitude $\eta_1$. Different colors correspond to a varying $\omega_{\mathrm{eff}}$. (b) Bosonic mode population for $\omega_{\mathrm{eff}}/2\pi=\SI{5}{MHz}$ and a varying drive amplitude $\eta_1$. The evolution and maximum population is confirmed to be independent of $\eta_1$ in first order. Master equation simulations here are performed without taking into account dissipation and neglect parasitic driving of the bosonic mode.}
\label{fig:clsim}
\end{centering}
\end{figure}

Since a rotating frame is not an inertial system, the laws of physics may be drastically altered when describing a physical system in a rotating frame. In the framework of analog quantum simulation, this offers a rich toolbox which allows to access intriguing effective parameter regimes that are hard or impossible to access in the laboratory frame. While the qubit dynamics in the rotating frame is well reproduced as demonstrated in Fig.~3 in the main text, it is not a priori clear that the bosonic mode population evolution of the quantum Rabi model is well reflected by the laboratory frame simulation. To verify the simulation scheme for the ideal system we compare the bosonic mode population in the driven laboratory frame, and the expected evolution of the ideal quantum Rabi model via classical master equation simulations. Here, we neglect dissipation and parasitic driving of the bosonic mode. \ref{fig:clsim}(a) demonstrates good agreement when comparing the time evolutions of the ideal (solid line) and the constructed (dashed line) Hamiltonians. The violation of excitation number conservation in the quantum Rabi model manifests in excitation numbers of the bosonic mode of larger than one for small $\omega_{\mathrm{eff}}$. We find that the maximum photon excitation in the bosonic mode roughly equals one for choosing simulation conditions where $\omega_{\mathrm{eff}}\approx g\sim 2\pi\times\SI{5}{MHz}$.

For $\omega_{\mathrm{eff}}/2\pi=\SI{5}{MHz}$ we demonstrate that the photon population in the bosonic mode is independent of the applied drive amplitude $\eta_1$, reflecting the fact that it does not appear in the synthesized Hamiltonian, Supplementary Equation~\eqref{eq:Hrot1eff}. Simulations for various $\eta_1$ are depicted in \ref{fig:clsim}(b).

\subsection{\label{sec:qretrieval}Qubit population retrieval from measured dispersive shifts of the readout resonator}

\begin{figure} 
\pagebreak
\includegraphics{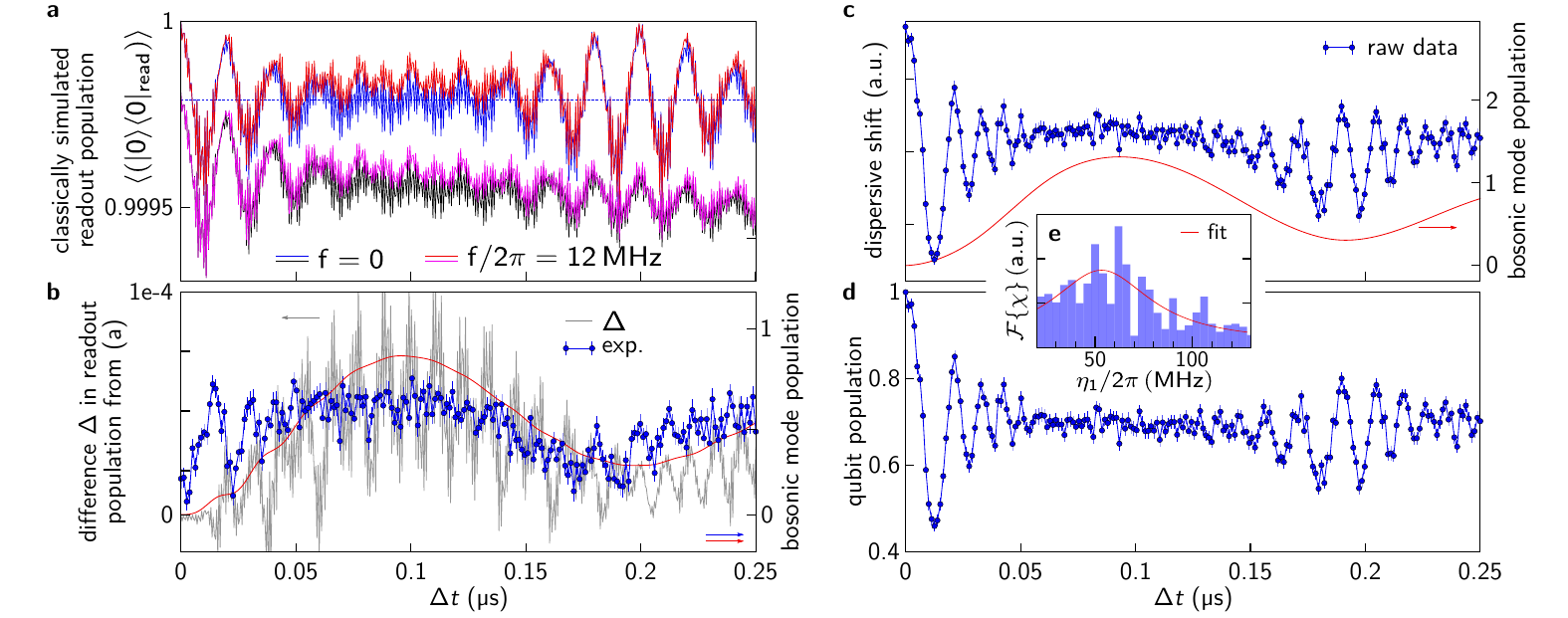}
\caption{\textbf{Verification of the photon exchange coupling between the bosonic mode and the readout resonator} (a) Classical master equation simulation of the vacuum projection $\Ket 0\Bra 0$ of the readout resonator state for vanishing photon exchange coupling $f =0$ (blue) and for $f/2\pi =\SI{12}{MHz}$ (red), disregarding dissipation. The blue dashed line denotes the mean value of the $f =0$ simulation as a guide to the eye. Magenta and black traces show the same classical simulations in the presence of dissipation and are shifted for better visibility. The additional bulge in the presence of the photon exchange coupling $f>0$ is apparent. (b) The difference (gray) between both classical simulations from (a) including dissipation follows the experimentally measured bosonic mode population (blue). Both data sets are fitted to the time evolution of the bosonic mode population (red), as obtained from the same classical simulation. (c) Dispersive shift of the readout resonator including both the shifts induced by qubit and bosonic mode. The red line depicts the classically simulated population of the bosonic mode. (d) Extracted qubit signal in the original qubit basis after subtracting the additional shift induced by the bosonic mode, based on the classically simulated expectation depicted in (c). The qubit was prepared in $\Ket{\mathrm{e}}$. From Fourier transformation (e) we obtain $\eta_1/2\pi\sim \SI{52}{MHz}$. Error bars throughout the figure denote a statistical standard deviation as detailed in the Methods of the main text.}
\label{fig:crossKerr}
\end{figure}

In measuring the dispersive shift of the readout resonator during the quantum simulation experiment, we observe a bulged and shifted equatorial baseline following the expected population evolution of the bosonic mode as obtained from the master equation simulation. We attribute this to a photon exchange coupling $f$ between the bosonic mode and the readout resonator, potentially mediated by the qubit. By inheriting nonlinearity from the qubit it gives rise to a dispersive shift on the readout resonator dependent on the photon number in the bosonic mode. The complete Hamiltonian including the readout resonator of resonance frequency $\omega_{\mathrm{r}}$ with creation (annihilation) operator $\hat a ^\dagger$ ($\hat a$) and with the RWA applied takes the form
\begin{align}
\frac{\hat H}{\hbar} =&\frac{\epsilon} 2 \hat\sigma _z+\omega\hat b ^\dagger \hat b +\omega_{\mathrm{r}} \hat a _{\mathrm{r}} ^\dagger \hat a _{\mathrm{r}} +g\left(\hat\sigma _-\hat b ^\dagger +\hat\sigma _+\hat b\right) +g_{\mathrm{r}}\left(\hat\sigma _-\hat a _{\mathrm{r}} ^\dagger +\hat\sigma _+\hat a _{\mathrm{r}}\right)\nonumber\\ &+f\left(\hat a _{\mathrm{r}}\hat b ^\dagger +\hat a _{\mathrm{r}} ^\dagger \hat b\right) +\hat\sigma _x\eta_1\cos(\omega_1 t+\varphi_1)+\hat\sigma _x \eta_2\cos(\omega_2 t+\varphi_2).
\label{eq:crossKerr}
\end{align}
$g_{\mathrm{r}}/2\pi\sim \SI{55}{MHz}$ denotes the coupling strength between qubit and readout resonator. The conjecture is verified by comparing classical master equation simulations with the photon exchange coupling $f$ switched on and off, respectively, see \ref{fig:crossKerr}(a). As visible in \ref{fig:crossKerr}(b), the difference of both classical simulations (gray) follows the evolution of the bosonic mode population (red) in the rotating frame as obtained from the same simulation. The experimentally extracted bosonic mode population (blue) likewise agrees with the trend from the classical simulation. The photon exchange coupling assumes the form of a cross-Kerr interaction after diagonalization in the subspace spanned by the bosonic mode and the readout resonator. We isolate this additional dispersive shift $\propto f^2$ by adding up measured data for the qubit prepared either in $\Ket{\mathrm{g}}$ or $\Ket{\mathrm{e}}$, as described in the Methods of the main text. Comparison with measured data suggests $f\sim\SI{}{MHz}$.

\subsection{\label{sec:sigxterm}Effect of the qubit tunneling term in the effective Hamiltonian}

\begin{figure}[H] 
\begin{centering}
\includegraphics{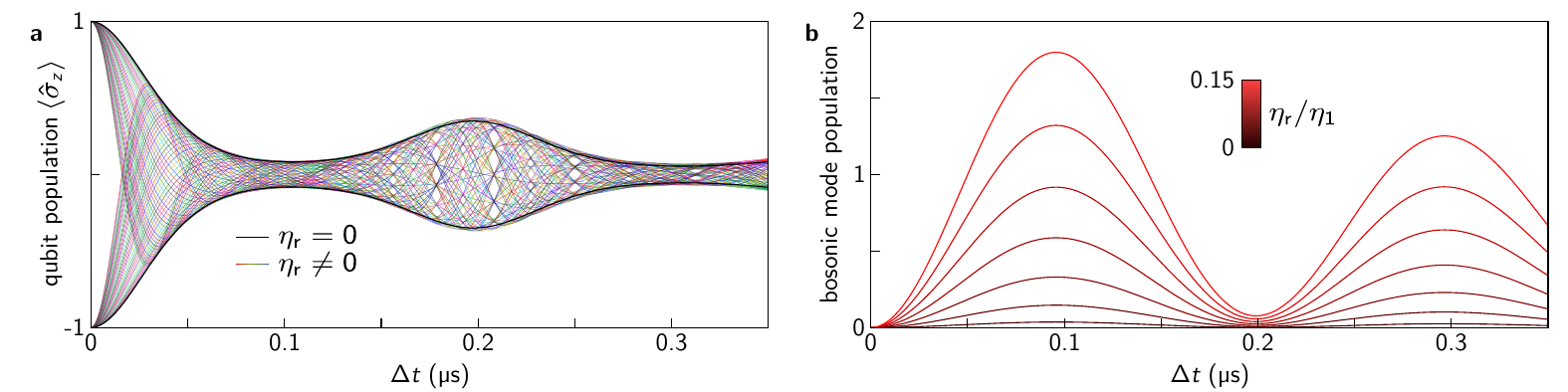}
\caption{\textbf{Qubit tunneling term in the effective Hamiltonian} (a) Classical simulation of the qubit population for the effective Hamiltonian, Eq.\eqref{eq:displHeff}, without the qubit tunneling term $\eta_{\mathrm{r}}=0$ (black) in comparison with simulations for $\eta_{\mathrm{r}}\neq 0$, plotted in colors. One can see that the additional qubit tunneling term $\propto \hat \sigma _x$ introduces a rotation while complying with the envelope of the ideal Hamiltonian, Supplementary Equation~\eqref{eq:Hrot1eff}. We demonstrate this by plotting the evolution for various values of $\eta_{\mathrm{r}}$ and $\omega_{\mathrm{eff}}/2\pi=\SI{5}{MHz}$. (b) Classically simulated bosonic mode population for a harmonic oscillator under static transversal drive. The periodicity of the evolution in the quantum Rabi model is reproduced up to a scaling factor.}
\label{fig:sigmax}
\end{centering}
\end{figure}

For a realistic parasitic coupling $\eta_{\mathrm{r}} \sim 0.1\eta_1$ of the dominant Rabi drive to the bosonic mode and $\omega_{\mathrm{eff}}/2\pi=\SI{5}{MHz}$, we obtain an effective qubit tunneling term
\begin{equation}
g\frac{\eta_{\mathrm{r}}}{2\omega_{\mathrm{eff}}}\hat \sigma _x \sim 2\pi\times \SI{2.2}{MHz}\times \hat \sigma _x \sim 0.4\omega_{\mathrm{eff}}\times\hat \sigma _x,
\end{equation}
by using the displaced effective Hamiltonian derived in Supplementary Equation~\eqref{eq:displHeff}. The thick black line in \ref{fig:sigmax}(a) shows a classical simulation of the qubit population according to the effective Hamiltonian, Supplementary Equation~\eqref{eq:Hrot1eff} without parasitic driving. For switching on the parasitic driving, $\eta_{\mathrm{r}}\neq 0$, the qubit is subject to a sub-rotation that however adheres with the envelope defined by the pure Hamiltonian. The dynamics of the bosonic mode is not qualitatively altered by the parasitic drive, as the displacement transformation defined in Supplementary Equation~\eqref{eq:displ}
\begin{equation}
\hat D ^\dagger \left(\omega_{\mathrm{eff}} \hat b ^\dagger \hat b +\frac 1 2 \eta_{\mathrm{r}} (\hat b ^\dagger +\hat b)\right)\hat D
= \omega_{\mathrm{eff}} \hat b ^\dagger \hat b +\mathrm{const.}
\end{equation}
leaves the eigenenergies of the isolated harmonic oscillator unchanged. The time evolution of the Hamiltonian $\omega_{\mathrm{eff}} \hat b ^\dagger \hat b +\frac 1 2 \eta_{\mathrm{r}} (\hat b ^\dagger +\hat b)$ is depicted in \ref{fig:sigmax}(b) for varying $\eta_{\mathrm{r}}$.


\section{\label{sec:revivals}Quantum revivals for initial qubit state $\Ket{\mathrm{g}}$, $\Ket{\mathrm{e}}$}

\begin{figure}[H] 
\begin{centering}
\includegraphics{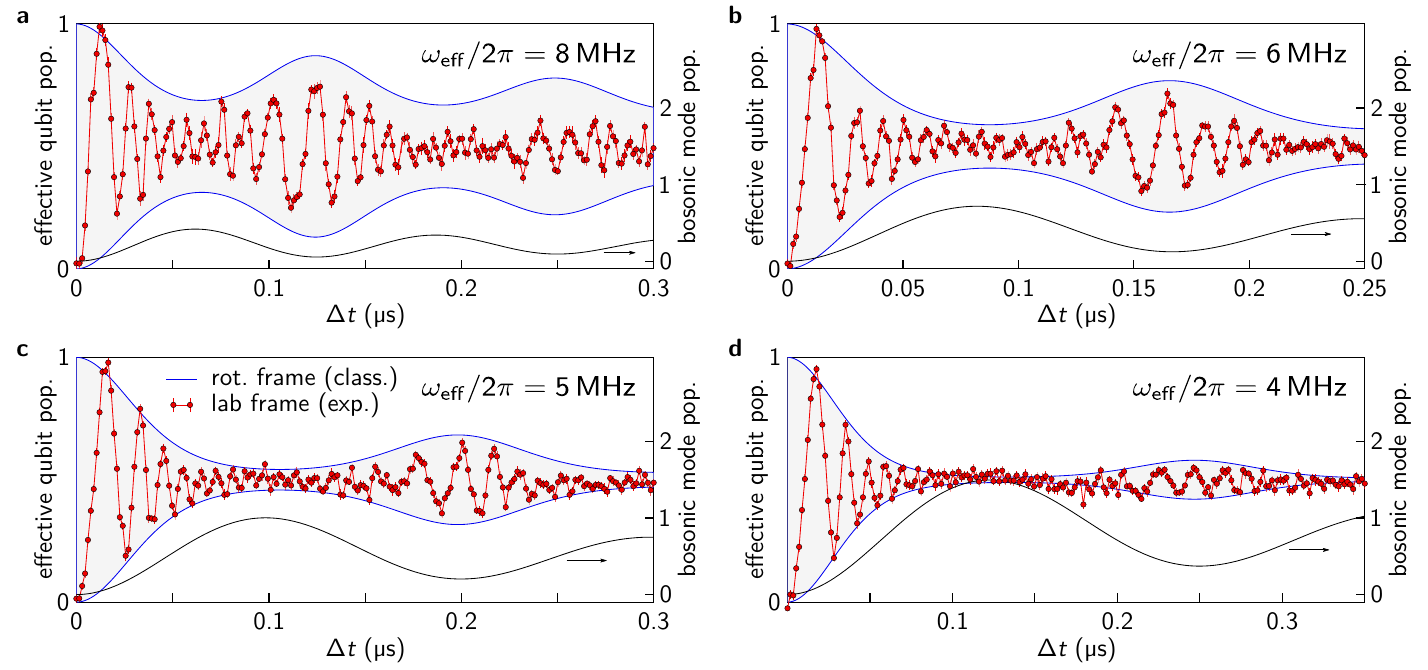}
\caption{\textbf{Quantum collapse and revival signatures for various $\omega_{\mathrm{eff}}$} The qubit is prepared in its groundstate $\Ket{\mathrm{g}}$ and the bosonic mode is initially in the vacuum state $\Ket 0$. The first revival appears at $2\pi/\omega_{\mathrm{eff}}$, respectively, and the different plots correspond to a varying $\omega_{\mathrm{eff}}$. The blue line shows a classical master equation simulation of the ideal effective Hamiltonian in the rotating frame, while the red data points are the qubit population evolution in the effective qubit frame. The black line shows the classically simulated bosonic mode population in the rotating frame, which increases with decreasing $\omega_{\mathrm{eff}}$. Note that the scale on the horizontal axis is not equal for each plot. The depicted qubit signal is extracted from measured data with the protocol described in the Methods in the main text and based on the classical simulations (black) in the absence of a parasitic drive of the bosonic mode. Error bars denote a statistical standard deviation as detailed in the Methods of the main text.}
\label{fig:revivals}
\end{centering}
\end{figure}

We demonstrate that the position of the quantum revival corresponds to $2\pi/\omega_{\mathrm{eff}}$ in quantum simulations with initial state prepared in $\Ket{0}\otimes\Ket{\mathrm{g}}$, where both the qubit and the bosonic mode are in its groundstate. Classical simulations are done for $\kappa\sim 3.9\times 10^ 6\,\mathrm{s^{-1}}$ and $1/T_1=0.2\times 10^ 6\,\mathrm{s^{-1}}$, $1/T_2=2.0\times 10^ 6\,\mathrm{s^{-1}}$ and in the absence of a parasitic drive of the bosonic mode.

We find a better agreement of experimental data with classical simulations for a slightly increased geometric coupling $g/2\pi\sim\SI{5.5}{MHz}$ as compared to the measured value in Fig.~2 in the main text of $g/2\pi=\SI{4.3}{MHz}$. This can be explained by the slight excess population due to the parasitic coupling of the Rabi drives to the bosonic mode and the effect may be additionally enhanced by a population of higher transmon levels. Increased decay and decoherence rates, accounting for a changed environmental spectral density in the rotating frame, did not lead to the correct ratio of revival and idling amplitudes. Simulations using $g/2\pi=\SI{4.3}{MHz}$ show results with a $\sim 30\%$ decrease in the maximum population of the bosonic mode with no qualitative consequences. With the measured geometric coupling we approach an USC regime with $g_{\mathrm{eff}}/\omega_{\mathrm{eff}}\sim 0.6$ for the experiment depicted in \ref{fig:revivals}(d). With the slightly increased value for $g_{\mathrm{eff}}$ we reach a relative coupling ratio of $0.7$. Measured values of $g_{\mathrm{eff}}$ and the effective value are summarized in \ref{tab:ggeo}.


\section{\label{sec:eta2comp}Verification of the parameter constraint for $\eta_2\neq 0$}

\begin{figure}[H] 
\begin{centering}
\includegraphics{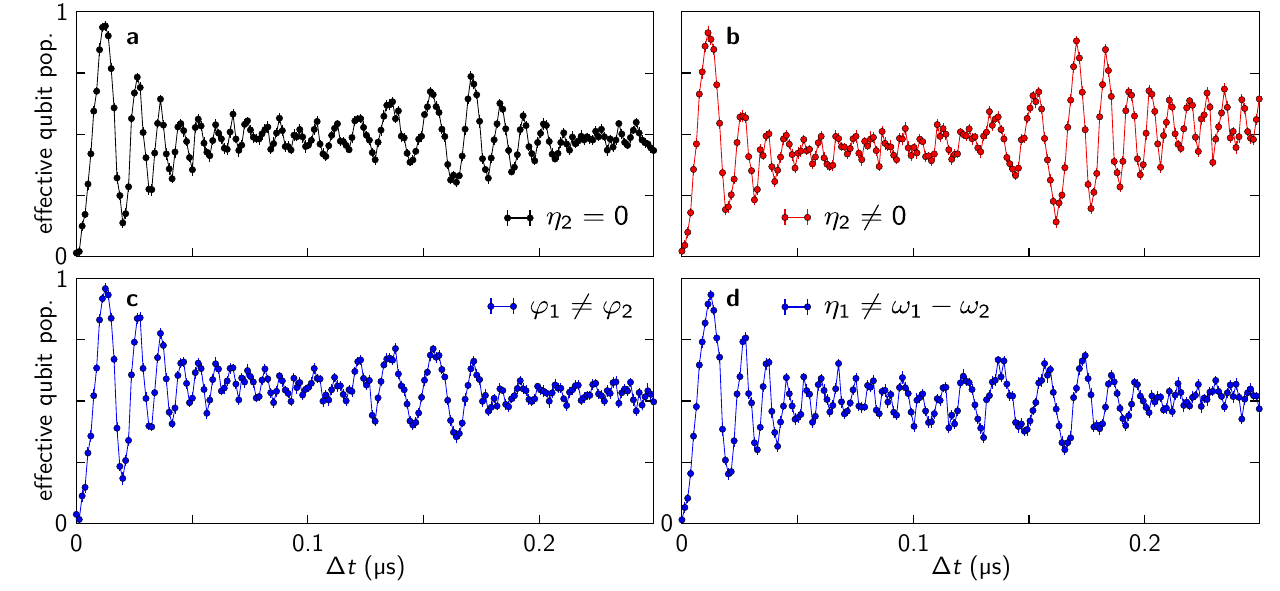}
\caption{\textbf{Verification of the parameter constraint for $\eta_2\neq 0$} Measurements for $\eta_2=0$ (a) and the proper parameter choice (b) when $\eta_2\neq 0$ as shown in Fig.~4 in the main text with $\eta_1/2\pi=\SI{58}{MHz}$. The measurement satisfying the required parameter constraints (b) is compared to measurements with parameters intentionally violating the phase matching condition, $\varphi_1\neq\varphi_2$ (c) and the condition $\eta_1\neq\omega_1-\omega_2$ (d). When the required constraints are not satisfied, the expected signatures are suppressed. The dispersive shift of the readout resonator induced by the bosonic mode is subtracted based on classically simulated data. Error bars throughout the figure denote a statistical standard deviation as detailed in the Methods of the main text.}
\label{fig:eta2comp}
\end{centering}
\end{figure}

We verify the simulation data for the full quantum Rabi model presented in the main text by comparison to measured data with intentionally departing from the required parameter constraints. While an increase in revival amplitude and an increased amplitude of the fast oscillations for larger $\Delta t$ is visible by comparing \ref{fig:eta2comp}(a), (b) presented in the main text, these signatures vanish for violating the parameter conditions
\begin{align}
\varphi_1=\,&\varphi_2\nonumber\\
\eta_1=\,&\omega_1-\omega_2
\end{align}
required by the simulation scheme. We chose $\varphi_1\sim\varphi_2+\pi$ but $\eta_1=\omega_1-\omega_2$ in \ref{fig:eta2comp}(c) and $\omega_2=\omega_1-\eta_1-2\pi\times\SI{10}{MHz}$ while $\varphi_1=\varphi_2$ in \ref{fig:eta2comp}(d) in order to demonstrate that the desired experimental features vanish.

An additional limitation of the simulation quality is imposed by an uncertainty of the effective Rabi frequency $\eta_1$, which is extracted from a rather broad peak in the Fourier transformed qubit evolution, see \ref{fig:crossKerr}. This causes the main constraint of the simulation scheme $\eta_1=\omega_1-\omega_2$ to be poorly satisfied in particular at small simulation times.


\section{\label{sec:spec}Spectroscopy of the avoided crossing between qubit and bosonic mode}

\begin{figure}[H] 
\begin{centering}
\includegraphics{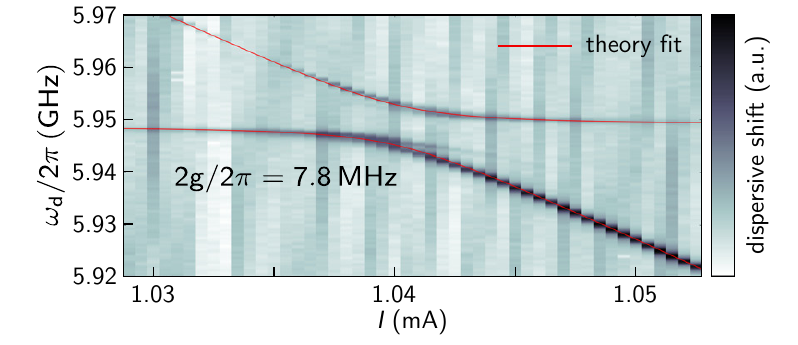}
\caption{\textbf{Avoided crossing between qubit and bosonic mode in spectroscopy} The qubit transition frequency is tuned by a dc current applied to the flux coil. The dispersive shift of the readout resonator is proportional to the excitation number of the qubit and is depicted in colors.}
\label{fig:spec}
\end{centering}
\end{figure}

The coupling between qubit and bosonic mode is pre-characterized in a two tone spectroscopy measurement using a vector network analyzer and a microwave source. The dispersive readout resonator shift is measured with a continuous microwave probe tone and an additional microwave drive tone is applied through the same transmission line to excite the qubit transition. The fit of the observed avoided crossing yields a minimum line separation of $2g/2\pi=\SI{7.8}{MHz}.$